\documentclass[prl,aps, superscriptaddress,twocolumn]{revtex4-1}
\usepackage{amsmath,amssymb}
\usepackage{graphicx}
\usepackage{wasysym}
\usepackage{amsfonts}
\usepackage{bm}
\usepackage{enumerate}
\usepackage[dvipsnames]{xcolor}
\usepackage[resetlabels]{multibib}
\usepackage{epstopdf}
\usepackage{latexsym}
\usepackage{multirow}
\usepackage{booktabs}
\usepackage[breaklinks,colorlinks = true,linkcolor = red,urlcolor=cyan,citecolor=red]{hyperref}
\usepackage[caption=false,singlelinecheck=false]{subfig}

\usepackage{times}
\newcommand{\bea}{\begin{eqnarray}}
\newcommand{\eea}{\end{eqnarray}}
\newcommand{\be}{\begin{eqnarray}}
\newcommand{\ee}{\end{eqnarray}}
\newcommand{\bw}{\begin{widetext}}
\newcommand{\ew}{\end{widetext}}
\newcommand{\nn}{\nonumber}

\newcommand{\la}{\langle}
\newcommand{\ra}{\rangle}

\makeatletter
\newcommand*{\sumcirclearrowleft}{%
  \DOTSB
  \mathop{
    \mathchoice
      {\rlap{\kern.25em\rotatebox[origin=c]{-90}{$\circlearrowleft$}}{\sum}}
      {\vcenter{\rlap{\kern.2em\rotatebox[origin=c]{-90}{$\scriptscriptstyle\circlearrowleft$}}}{\sum}}
      {\sum}{\sum}
  }\slimits@
}

\newcommand*{\sumcirclearrowright}{%
  \DOTSB
  \mathop{
    \mathchoice
      {\rlap{\kern.25em\rotatebox[origin=c]{90}{$\circlearrowright$}}{\sum}}
      {\vcenter{\rlap{\kern.2em\rotatebox[origin=c]{90}{$\scriptscriptstyle\circlearrowright$}}}{\sum}}
      {\sum}{\sum}
  }\slimits@
}
\makeatother

\usepackage{soul}

\makeatletter
\def\SOUL@ulstunderline#1{{%
    \setbox\z@\hbox{#1}%
    \dimen@=\wd\z@
    \dimen@i=\SOUL@uloverlap
    \advance\dimen@2\dimen@i
    \rlap{
        \null
        \kern-\dimen@i
        \SOUL@ulcolor{\SOUL@ulleaders\hskip\dimen@}%
    }%
    \SOUL@stpreamble
    \rlap{%
        \null
        \kern-\dimen@i
        \SOUL@ulcolor{\SOUL@ulleaders\hskip\dimen@}%
    }%
    \unhcopy\z@
}}
\def\SOUL@ulsteverysyllable{%
    \SOUL@ulstunderline{%
        \the\SOUL@syllable
        \SOUL@setkern\SOUL@charkern
    }%
}
\def\SOUL@ulstsetup{%
  \SOUL@ulsetup
  \let\SOUL@everysyllable\SOUL@ulsteverysyllable
}
\DeclareRobustCommand*\textulst{\SOUL@ulstsetup\SOUL@}
\makeatletter

\begin{document}

\title{Hidden phases born of a quantum spin liquid: Application to pyrochlore spin ice}

\author{Hyeok-Jun Yang}
\email{yang267814@kaist.ac.kr}
\affiliation{Department of Physics, Korea Advanced Institute of Science and Technology, Daejeon, 34141, Korea}

\author{Nic Shannon}
\email{nic.shannon@oist.jp}
\affiliation{Theory of Quantum Matter Unit, Okinawa Institute of Science and Technology Graduate University,Onna-son, Okinawa 904-0412, Japan}

\author{SungBin Lee}
\email{sungbin@kaist.ac.kr}
\affiliation{Department of Physics, Korea Advanced Institute of Science and Technology, Daejeon, 34141, Korea}
\date{\today}

\begin{abstract}

Quantum spin liquids (QSL) have 
generated considerable excitement 
as phases of matter with emergent gauge structures and fractionalized excitations.
In this context, phase transitions out of QSLs have been widely discussed as Higgs 
transitions from deconfined to confined phases of a lattice gauge theory.
However the possibility of a wider range of novel phases, occuring between 
these two limits, has yet to be systematically explored.
In this Letter, we develop a formalism which allows for interactions between fractionalised 
quasiparticles coming from the constraint on the physical Hilbert space, and can be used to 
search for exotic, hidden phases.
Taking pyrochlore spin ice as a starting point, we show how a U(1) QSL can give 
birth to abundant daughter phases, without need for fine--tuning of parameters.
These include a (charged--) $\mathbb{Z}_2$ QSL, and a supersolid.
We discuss implications for experiment, and numerical results which support our analysis.  
These results are of broad relevance to QSL subject to a parton description, 
and offer a new perspective for searching exotic hidden phases in quantum magnets.

\end{abstract}

\maketitle


\label{sec:Introduction}
{\textbf{\textit {  Introduction ---}}}
One of the most intriguing features of frustration in magnets is the possibility of finding 
phases of matter which lie outside the usual Landau paradigm of symmetry--broken states
\citep{Anderson1973, Fazekas1974}. 
A prominent example is the ``quantum spin liquid'' (QSL), 
a state where spins continue to fluctuate at low temperatures, 
achieving a massively--entangled 
state with fractionalised excitations \citep{Balents2010, Gingras2014, Savary2016}. 
Long a subject of theoretical conjecture, in the past decade QSL have also 
become an intense focus for experimental research \cite{Lee2008,Zhou2017,Knolle2019}.
One of their defining properties is non--locality, frequently encoded in an 
emergent gauge degree of freedom.
And in this respect, the parton approach has proved a powerful tool for describing 
both the fractionalised excitations of QSL, and their (non--local) 
interactions \citep{Wen2004-book}. 


Besides being interesting in their own right, QSL also give rise to  
``daughter'' phases 
which may have an unconventional, or hidden, character.  
QSL described by the deconfined phase of a lattice gauge theory 
\citep{Wegner1971, Wilson1974, Kogut1979, Fradkin1979}, 
prove surprisingly stable against weak perturbations 
\citep{Senthil2000, Moessner2003, Hermele2004, Montrunich2005, Fradkin1990, Banerjee2008, McClarty2015}.
None the less, at strong coupling, they can undergo confinement through a 
Higgs transition \citep{Anderson1963,Higgs1964}, 
into a magnetically--ordered phase which breaks the emergent gauge symmetry 
\citep{Fradkin1979, Banerjee2008, Powell2011}.
What happens at intermediate coupling, where usual 
perturbation theory breaks down, remains an open question.
In particular, the possibility of finding new, intermediate phases 
between the QSL and Higgs phases at strong coupling, has 
yet to be systematically explored.
%


In this Letter, we argue that novel phases, intermediate between the 
confined and deconfined limits of a pure lattice gauge theory, may be a 
generic feature of models supporting QSL.
These intermediate phases are driven by interactions between the collective 
excitations of the QSL, which are obscured in the perturbative limit of the problem.   
In particular, fluctuations of the (gauge--)charge, absent in a pure gauge theory, 
generate new effective interactions between the fractionalised excitations of the QSL.
As a result, the lifting of gauge symmetry can become a two--step process, 
with a new intermediate phase, also with QSL character, occurring between 
the orginal QSL and its fully--confined, Higgs phase.


We develop these ideas in the context of a parton theory of pyrochlore 
quantum spin ice, where the existence of a  deconfined $U(1)$ QSL and 
its corresponding, magnetically--ordered Higgs phase, are already well established.
Starting from a standard Bosonic parton prescription, we develop a formalism 
which takes into account the fact that charge fluctuations are bounded
by the finite Hilbert space of the underlying spins.
Applying this to an extended XXZ model, we find that the 
$U(1)$ QSL can give rise to a plethora of exotic 
phases, including a $\mathbb{Z}_2$ QSL; 
a ``charged'' $\mathbb{Z}_2$ QSL with broken inversion symmetry;  
and a ``spinon supersolid'' which breaks both inversion and time--reversal 
symmetries.
Possible experimental signatures of these phases are discussed.
While this particular hierarchy of phases is specifc to the pyrochlore lattice, 
the formalism developed is quite general, and should help to guide 
the search for hidden phases in a wide range of quantum magnets.


\label{sec:Pyrochlore spin ice}
%
{\textbf{\textit {  Pyrochlore spin ice ---}}}
Pyrochlore oxide materials with a chemical formula $\text{R}_2\text{TM}_2\text{O}_7$ (\text{R}: 
rare earth, \text{TM} transition metal) \citep{Gardner1999, Cao2009} 
%
%
have proved a rich source of candidates for 
QSL and related forms of order
\cite{Ross2011, Thompson2011,Chang2012,Kimura2013,Wen2017,Sibille2018,Gaudet2019,Sibille2020,Xu2020}.
In many of these materials, localized f--electrons form a (non--)magnetic doublet 
described by a pseudospin--1/2, 
with the minimal model taking the form \citep{Onoda2011, Onoda_2011}  
%
%
\begin{eqnarray}
H_{\text{XXZ}}&&=H_0+H_1
\nonumber\\
&&=\sum_{\langle ij\rangle} \Big[ J_z S_i^zS_{j}^z - J_{\pm} (S_i^+S_{j}^- + \text{h.c.})\Big] \; ,
\label{eq:H.XXZ}
\end{eqnarray}
where the sum $\langle ij\rangle$ runs over the first--neighbour bonds 
of a pyrochlore lattice.
In the perturbative limit $J_z \gg |J_\pm|$, $(J_z>0)$, the dominant Ising term $H_0$ favors 
an extensively--degenerate set of classical spin ice states \cite{Anderson1956,Harris1997, Bramwell2001}, while the spin--flip term $H_1$ causes mixing of these states, leading 
to a QSL ground state \citep{Hermele2004, Banerjee2008, Shannon2012, Benton2012, Kato2015,McClarty2015,Huang2018,Huang2020}.   
%


This spin liquid can be elegantly described in terms of a compact, 
frustrated, $U(1)$ lattice gauge theory \cite{Hermele2004,Benton2012,Savary2012,Lee2012}.
This is defined on the sites $\textbf{r}, \textbf{r}'$ of a (bipartite) 
diamond lattice,  
with spin operators expressed in terms of an 
emergent gauge field $A_{\textbf{r}\textbf{r}'}\; (\text{mod}\; 2\pi)$, 
electric field $E_{\textbf{r}\textbf{r}'}$ (half--integer), and 
matter field $\phi_\textbf{r}=e^{-i\varphi_\textbf{r}}$ (a Bosonic spinon),  
conjugate to a gauge charge $Q_{\textbf{r}}$, such that
\begin{eqnarray}
&&Q_{\textbf{r}}=\eta_\textbf{r}\sum_{\mu=0}^3S_{\textbf{r},\textbf{r}+\eta_\textbf{r} \textbf{e}_\mu}^z, \quad S_{\textbf{r}\textbf{r}'}^z=\eta_\textbf{r} E_{\textbf{r}\textbf{r}'},
\nonumber\\
&& S_{\textbf{r}\textbf{r}'}^+ =\phi_\textbf{r}^{\dagger} s_{\textbf{r}\textbf{r}'}^+ \phi_{\textbf{r}'} \;\; (s_{\textbf{r}\textbf{r}'}^+=\frac{1}{2}e^{iA_{\textbf{r}\textbf{r}'}}\;\; \text{for}\; \textbf{r}\in A),
\label{eq:OriRotor}
\end{eqnarray}
where $\eta_{\textbf{r}\in A(B)}=1(-1)$ distinguishes 
sites belong to the $A(B)$ sublattice.
For an $S=1/2$ (pseudo--)spin doublet, the gauge charge 
takes on integer values 
\begin{eqnarray}
   Q_r = 0, \pm 1, \pm 2 \ldots \pm 4S \; ,
\label{eq:constraint}
\end{eqnarray}
with spin fluctuations acting as ladder operators 
for this tower of states. 
[See supplementary material for details].

%
%
%

\begin{figure}[t]
  \begin{center}
  \includegraphics[width=0.6\linewidth]{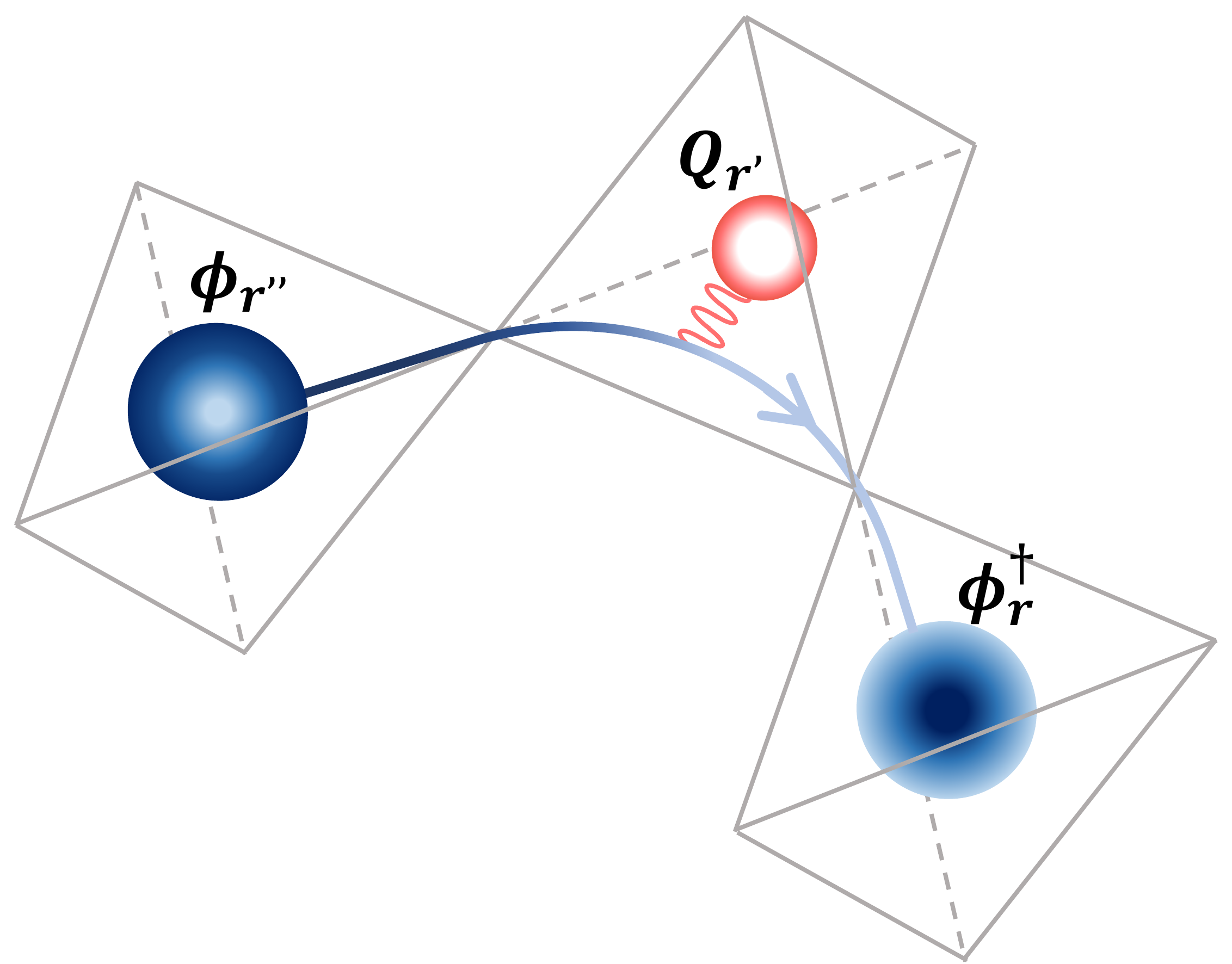}
     \caption{
     Schematic illustration of spinon propagation, showing how hopping of the spinon from 
     the diamond lattice site ${\bf r''}$ to site ${\bf r}$ couples to fluctuations of charge on 
     site ${\bf r'}$ [Eq.~(\ref{eq:H.rotor.extended})], leading to the effective spinon interaction 
     ${\mathcal H}_u$ [Eq.~(\ref{eq:H.u})].
     }    
     \label{fig:SpinonInt}
  \end{center}
\end{figure}

Following a standard prescrption \citep{Savary2012, Lee2012}, 
we further map the $U(1)$ lattice gauge theory onto a quantum rotor model 
of spinons coupled to a (static) $U(1)$ gauge field.
In doing so we consider the length of the spin $S$ 
in Eq.~(\ref{eq:constraint}) 
to be a formal control parameter, initially 
taking the limit $S \to \infty$.
Within these approximations
\begin{eqnarray}
&&\mathcal{Z}=\int \mathcal{D}\phi_\textbf{r}^*\mathcal{D}\phi_\textbf{r} \mathcal{D}Q_\textbf{r}|_{Q_\textbf{r}\in (-\infty,\infty)} e^{-\mathcal{S}_{\text{eff}}[\phi_\textbf{r}^*,\phi_\textbf{r},Q_\textbf{r}]} \; , 
\label{eq:Z.rotor}
\\
&&S_{\text{eff}}=\int_0^{\beta} d\tau \Big(\sum_\textbf{r}(iQ_\textbf{r}\partial_{\tau}\varphi_\textbf{r} +\lambda(\phi_\textbf{r}^*\phi_\textbf{r}-1)) + \mathcal{H}_{\text{rotor}}\Big) \; , \nonumber\\
\label{eq:S.rotor}
\end{eqnarray}
where the Lagrange multiplier $\lambda$ enforces the rotor constraint $|\phi_\textbf{r}|=1$ 
in a soft manner $\frac{1}{N}\sum_{\textbf{r}\in A} \langle \phi_\textbf{r}^{\dagger}\phi_\textbf{r} \rangle=1$, 
and 
\begin{eqnarray}
&&\mathcal{H}_{\text{rotor}} = \frac{J_z}{2}\sum_\textbf{r}  Q_\textbf{r}^2 
- \frac{J_\pm}{4}\sum_{\langle\langle\textbf{r}\textbf{r}''\rangle\rangle} \phi_{\textbf{r}}^{\dagger}e^{i( A_{\textbf{r}\textbf{r}'}+A_{\textbf{r}'\textbf{r}''})} \phi_{\textbf{r}''} +\text{h.c.} \; , \nonumber\\
\label{eq:H.rotor}
\end{eqnarray}
with spinons constrained to move on either the $A$ or $B$ sublattice of sites
$\langle\langle\textbf{r}\textbf{r}''\rangle\rangle$, 
cf. Fig.~\ref{fig:SpinonInt}.



Within the framework of this rotor model, 
transitions out of $U(1)$ QSL occur through the Higgs mechanism, 
and are associated with the Bose--Einstein condensation (BEC) of 
either electric charges (spinons) \citep{Gingras2014, Savary2016}, 
or the corresponding, dual, magnetic monopole \cite{Chen2016}.
Condensation of spinons leads to confined states with easy--plane 
magnetic order, and QMC simulations of Eq.~(\ref{eq:H.XXZ}) find 
easy--plane magnetic order with ${\bf q} =0$ for $J_\pm/J_z \gtrsim 0.05$, 
confirming the expected Higgs phase as the strong--coupling ground state \cite{Banerjee2008,Kato2015,Huang2020}.
A number of other forms easy--plane order have also been 
identified as Higgs phases in generic models of pyrochlore 
magnets \citep{Savary2012, Lee2012}.
However, to explore the possibility of new phases at intermediate 
coupling, a more general approach 
is needed.


Heuristically, we reason as follows:
In the limit \mbox{$J_\pm/J_z \to 0$}, fluctuations of the charge $Q_\textbf{r} \approx 0$ 
are neglible, and have no effect on the propagation of spinons.
However as $J_\pm$ increases, spinons begin to interact with a dilute cloud of 
charge fluctuations, which modify their dynamics, cf. Fig.~\ref{fig:SpinonInt}.
The simplest, gauge--invariant form of interaction capturing this effect is
\begin{eqnarray}
 \delta \mathcal{H}_{\text{rotor}} = g\ Q_{\textbf{r}'}^2\phi_\textbf{r}^{\dagger}e^{i(A_{\textbf{r}\textbf{r}'}+A_{\textbf{r}'\textbf{r}''})}\phi_{\textbf{r}''}+O(Q_\textbf{r}^4) \; ,
\label{eq:delta.H.rotor}
\end{eqnarray}
where the coupling constant $g$ increases with $|J_\pm|$.   
Physically, this describes the interplay of two competing tendencies, charge 
fluctuations, mediated by the motion of spinons, and the constraint on the 
maximum charge on a single site, reflected in the ladder termination
\bea
	S_{\textbf{r}\textbf{r}'}^+ |S_{\textbf{r}\textbf{r}'}^z =  S \rangle 
	= S_{\textbf{r}\textbf{r}'}^- |S_{\textbf{r}\textbf{r}'}^z = - S \rangle 
	= 0 \; .
	\label{eq:ladder.termination}
\eea
And it is this competition, absent in the usual rotor formulation, 
Eq.~(\ref{eq:H.rotor}), which has the potential to change the nature of the QSL.

%



\label{sec:Projected rotor representation}
{\textbf{\textit {  Projected rotor representation ---}}} 
Within the canonical rotor formalism, gauge 
charge takes on all real values \mbox{$-\infty < Q_\textbf{r} < \infty$} [Eq.~(\ref{eq:Z.rotor})].  
The consequences of the physical constraint $|Q_\textbf{r}| \leq 4S$ [Eq.~(\ref{eq:constraint})], 
can be explored through the new terms generated by restriction 
to a physical Hilbert space, viz
\begin{eqnarray}
\mathcal{Z}'&&=\int \mathcal{D}\phi_\textbf{r}^*\mathcal{D}\phi_\textbf{r} \mathcal{D}Q_\textbf{r}|_{Q_\textbf{r}\in (-4S, 4S)} e^{-\mathcal{S}_{\text{eff}}}
\nonumber\\
&&\approx\int \mathcal{D}\phi_\textbf{r}^*\mathcal{D}\phi_\textbf{r} \mathcal{D}Q_\textbf{r}|_{Q_\textbf{r}\in (-\infty, \infty)} e^{-\mathcal{S}_{\text{eff}}'} \; ,
\label{eq:S'}
\end{eqnarray}
where the effective action $\mathcal{S}_{\text{eff}}'$ [cf.~Eq.~(\ref{eq:S.rotor})], is modified 
through the projection of spinon and charge fields within the rotor Hamiltonian 
\begin{eqnarray}
\mathcal{H}_{\text{rotor}}'&&=\mathcal{H}_{\text{rotor}}[\tilde{Q}_\textbf{r},\tilde{\phi}_\textbf{r},\tilde{\phi}_\textbf{r}^{\dagger}; A_{\textbf{r}\textbf{r}'}, E_{\textbf{r}\textbf{r}'}] \quad(\tilde{O}_\textbf{r}\equiv P_\textbf{r} O_\textbf{r} P_\textbf{r}) \; . \quad \;
\label{eq:H.rotor.projected}
\end{eqnarray}
We require that the projection operator $P_\textbf{r}$ satisfies 
\begin{eqnarray}
&& P_\textbf{r}[Q_\textbf{r}] = 1\ \forall\ Q_\textbf{r} \in \{0,\pm 1,...,\pm 4S \} \; , \\
&& P_\textbf{r}|Q_\textbf{r} = \pm (4S+1)\rangle=0 \; .
\label{eq:projector}
\end{eqnarray}
ensuring that no matrix element of Eq.~(\ref{eq:H.rotor.projected}) connects to an 
unphysical state.
Resolving $P_\textbf{r}$ as a polynomial 
and expanding to leading order 
\begin{eqnarray}
P_\textbf{r}[Q_\textbf{r}]&&\equiv 1-\frac{1}{(4S+1)(8S+1)!}\prod_{k=0}^{4S}(Q_\textbf{r}^2-k^2)
\nonumber\\
&& =1-\alpha_S Q_\textbf{r}^2+\cdots \;\; \Big(\alpha_S=\frac{((4S)!)^2}{(4S+1)(8S+1)!}\Big) ,\quad\;\;\;
\label{eq:polynomial}
\end{eqnarray}
we find,  
\begin{eqnarray}
\mathcal{H}_{\text{rotor}}' = \mathcal{H}_{\text{rotor}} + \delta \mathcal{H}_{\text{rotor}} \; ,
\label{eq:H.rotor.extended}
\end{eqnarray}
where $\delta \mathcal{H}_{\text{rotor}}$ is given by Eq.~(\ref{eq:delta.H.rotor}), 
with $g = \alpha_S J_\pm$.   
Finally, integrating out $Q_\textbf{r}$ in Eq.~(\ref{eq:H.rotor.extended}), 
we arrive at an effective model with interaction between 
spinons
\begin{eqnarray}
&\mathcal{H}_\text{rotor}^{''} = \mathcal{H}_{\text{rotor}} + \mathcal{H}_u \; ,
\label{eq:H.interacting} \\
&\mathcal{H}_u = -u\sum_{\langle\langle \textbf{r}\textbf{r}''\rangle\rangle}\Big(\phi_\textbf{r}^{\dagger}e^{i(A_{\textbf{r}\textbf{r}'} + A_{\textbf{r}'\textbf{r}''})}\phi_{\textbf{r}''}\Big)^2+\text{h.c.} + \cdots , \quad \;\;
\label{eq:H.u} \\
&u  \sim  \alpha_S J_\pm^2/J_z \; .
\label{eq:parameter.u} 
\end{eqnarray}
We note that an approach based on cannonical transformations \cite{MacDonald1988} 
also leads to an attractive interaction between spinons, with the same form of vertex, 
at the same order in $J_\pm/J_z$. 
The role of longer--range interactions, omitted in Eq.~(\ref{eq:H.u}) 
will be discussed below.
[See supplementary material for details].


%
The projection method developed above has much in common with the expansion 
of interactions in spin--wave theory.
And just as in that case, where practioners must chose between the prescriptions of 
Hosltein and Primakoff \cite{Holstein1940}, and those of Dyson and 
Maleev \cite{Dyson1956,Maleev1958}, the form of projection, Eq.~(\ref{eq:polynomial}), 
is not uniquely determined.
None the less, since the structure of the vertex in Eq.~(\ref{eq:H.u}) is 
constrained by symmetry, it is not sensitive to the precise choice of projection operator.
With this in mind, we now proceed to examine the consequences of 
interactions between spinons.


\label{sec:Spinon interaction and hidden phases}
{\textbf{\textit {Spinon interaction and hidden phases ---}}} 
For large $u$, the implication of 
$H_u$ [Eq.~(\ref{eq:H.u})], can be understood by direct
analogy with earlier work on QSL described by a quantum 
rotor model \citep{Senthil2000}.
Here, since $u > 0$, the interaction mediates pairing between 
spinons, favouring a charge-2 condensate 
$\Delta\equiv \langle \phi_\textbf{r}\phi_\textbf{r}\rangle \neq 0$, 
which minimizes both terms in $\mathcal{H}_\text{rotor}^{''}$ [Eq.~(\ref{eq:H.interacting})].
Since this pairing occurs in the absence of a single--spinon condensate, 
$\langle \phi_\textbf{r} \rangle=0$, the state retains its fractionalised, 
QSL character, but with the gauge group broken from $U(1)$ down to  
$\mathbb{Z}_2$.
From Eq.~(\ref{eq:parameter.u}), we see that this new \mbox{``large--$u$''} 
phase is most likely to be realised for intermediate to large $J_\pm/J_z$, 
and for small values of $S$, i.e. in the physical limit of $S=1/2$.


\begin{figure}[t!]
\subfloat[]{\label{fig:schematic}\includegraphics[width=0.36\textwidth]{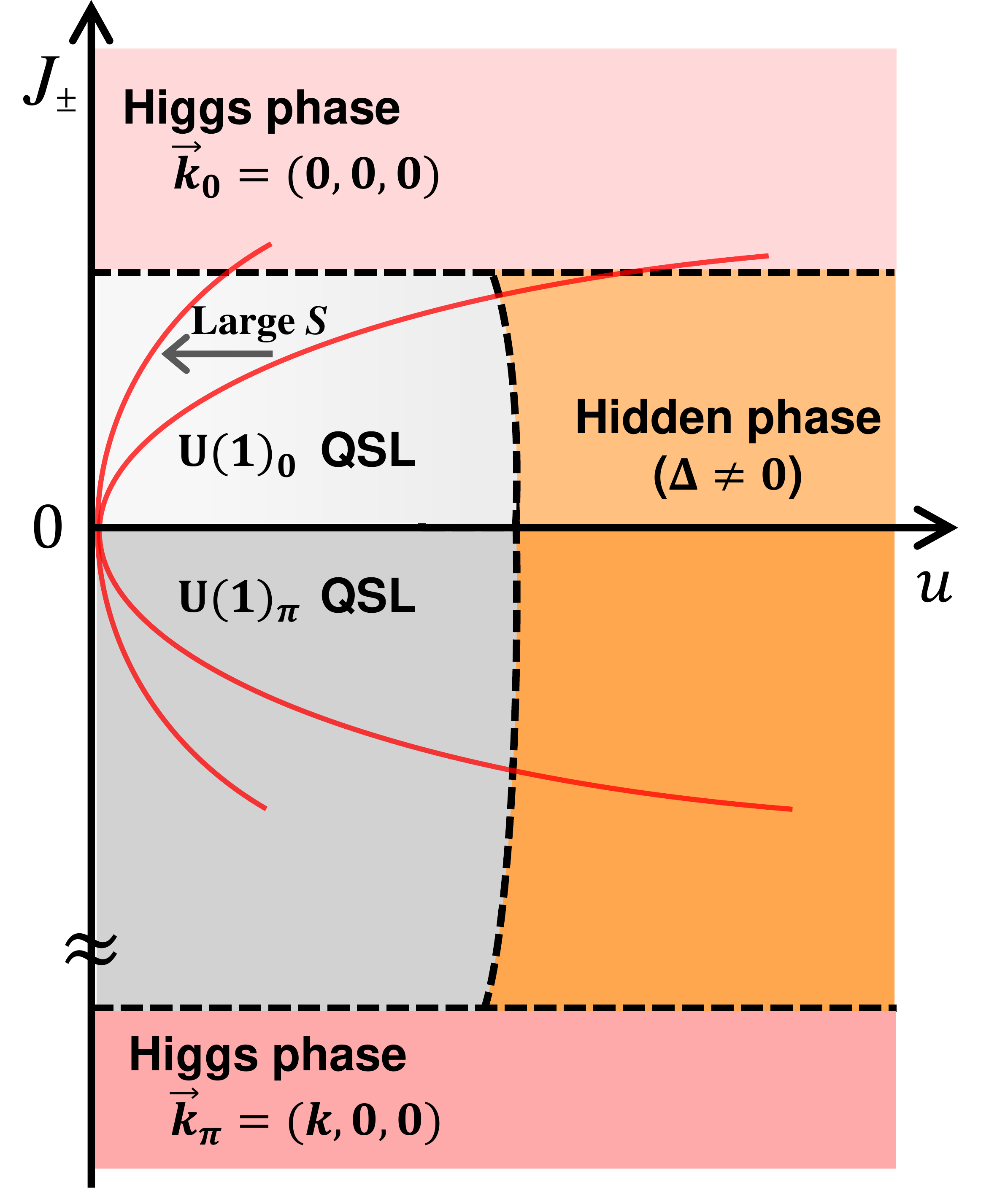}}
\subfloat[]{\label{fig:u=0.04}\includegraphics[width=0.12\textwidth]{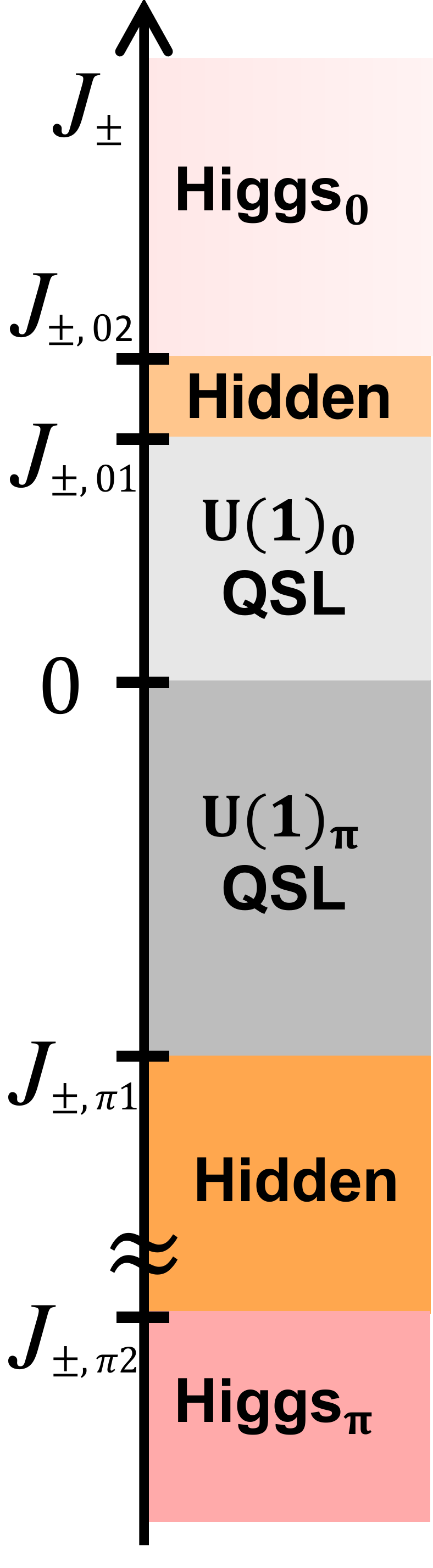}}
\caption{
\textul{Emergence of an intermediate, ``hidden'' phase in a model of quantum spin ice.}
(a) Schematic phase diagram of projected rotor model, $\mathcal{H}_\text{rotor}^{''}$ 
[Eq.~(\ref{eq:H.interacting})], as a function of transverse exchange $J_\pm$, and effective 
spinon interaction $u$ [Eq.~(\ref{eq:parameter.u})].
The ground state of the corresponding XXZ model $\mathcal{H}_\text{XXZ}$ 
[Eq.~(\ref{eq:H.XXZ})], evolves along a trajectory with curvature 
$\propto J_\pm^2/S$ (solid red line).
Where fluctuations of charge are significant (small $S$), this intersects 
a ``hidden'' phase intermediate between the QSL and ordered, Higgs phases.
The precise nature of each phase depends on the sign of $J_\pm$, 
as described in the text.
(b) Resulting phase diagram, as found within gauge mean-field theory 
(gMFT) \citep{Savary2012, Lee2012} for $S=1/2$.
Within this approximation, all phase transitions are 2$^{nd}$--order in character, 
with phase boundaries at $J_{\pm ,01} \approx 0.12$, $J_{\pm , 02} \approx 0.19$ 
and $J_{\pm ,\pi 1}\approx -0.26, J_{\pm ,\pi 2}\approx -4.13$, for $u=0.04$, 
in units of $J_z$. 
}
\label{fig:schematicdiagram}
\end{figure}


In Fig.~\ref{fig:schematic} we show a schematic phase diagram for the extended 
rotor model, Eq.~(\ref{eq:H.rotor.extended}), based on these expectations.
Known results for the rotor model, Eq.~(\ref{eq:H.rotor}),  are shown for $u=0$; 
here $U(1)$ QSLs with flux $0(\pi)$ give way to Higgs phases with ordering 
wave vectors $\textbf{k}_{0 (\pi)}$.
For $J_\pm \ll J_z$, these QSL become unstable at large $u$ against a 
``hidden'' phase with finite spinon pairing, $\Delta \ne 0$.
The behaviour expected of the XXZ model, Eq.~(\ref{eq:H.XXZ}), is shown 
through red parabolae with $u \sim J_\pm^2/S$.
Where fluctuations of charge are small (large $S$), 
the system passes directly from the $U(1)$ QSL to its assocated Higgs phase.
However where charge fluctuations are significant (small $S$) this trajectory 
can pass through the ``hidden'' phase, giving a range of $J_\pm/J_z$ for 
which the ground state is a $\mathbb{Z}_2$ QSL.
Specific estimates of the phase boundaries for $S=1/2$ and $u = 0.04\ J_z$ 
are shown in Fig.~\ref{fig:u=0.04}.

\begin{figure}[t]
\centering
  \subfloat[]{\label{fig:QQ}	\includegraphics[width=0.36\columnwidth]{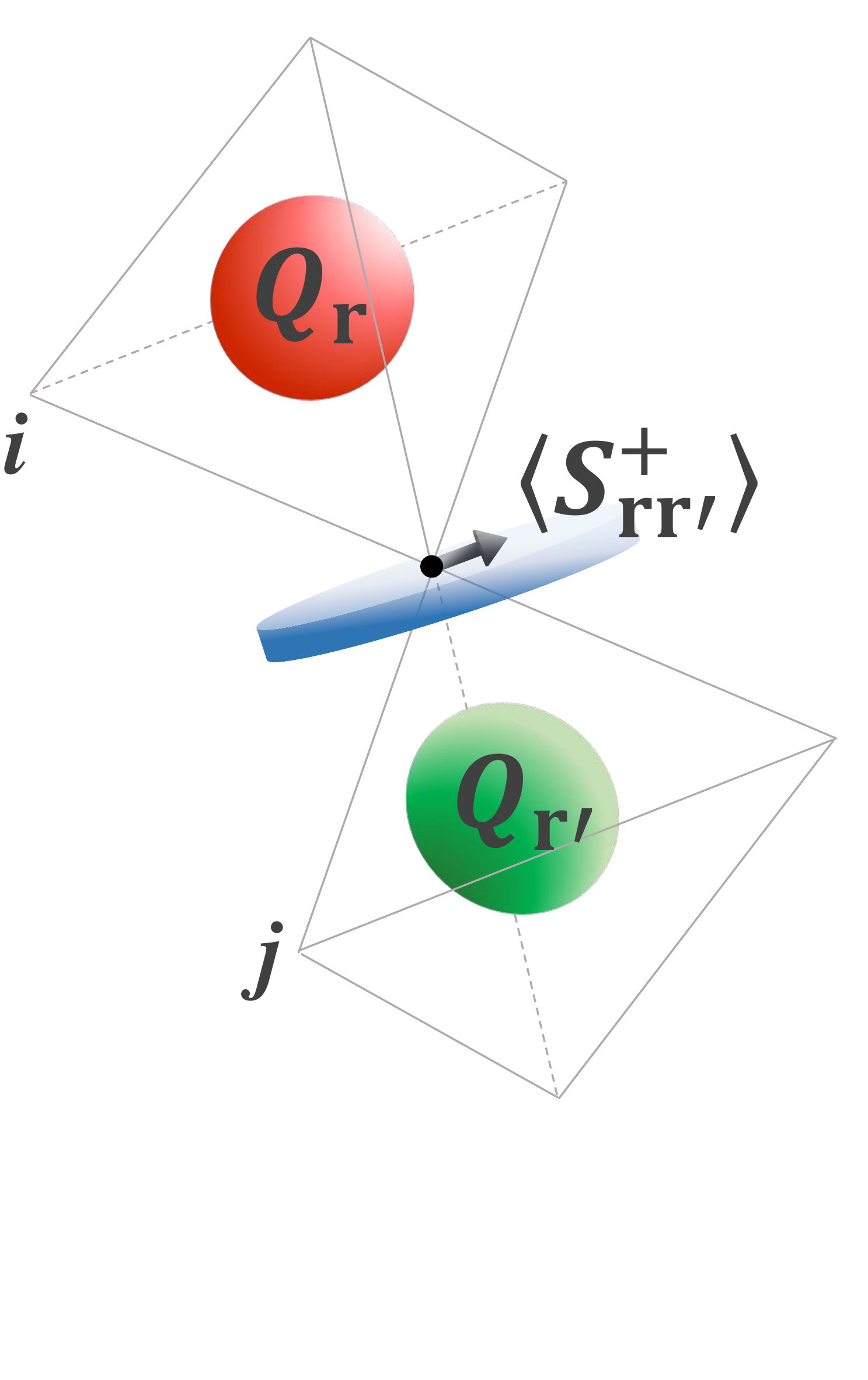}}
	\subfloat[]{\label{fig:ChargeInt}\includegraphics[width=0.64\columnwidth]{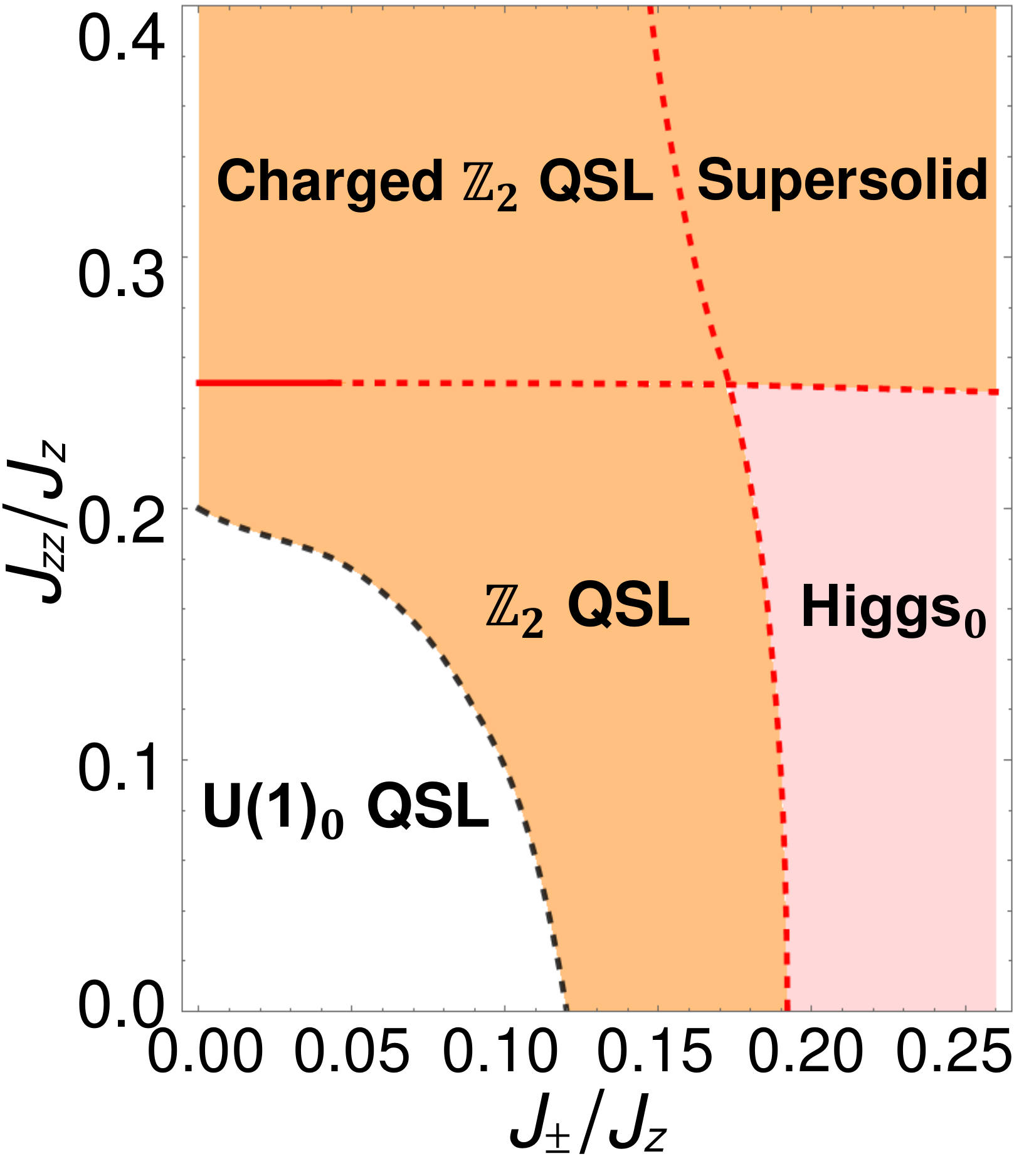}}
\caption{
\textul{Phase diagram of extended XXZ model,  
illustrating the possibility of charge--ordered phases.} 
(a) Convention for labelling sites in further--neighbor Ising interaction 
$\mathcal{H}_{\text{ZZ}}$ [Eq.~(\ref{eq:H.ZZ})].
Charge polarization on diamond--lattice sites $\langle Q_\textbf{r} \rangle$, 
can coexist with a transverse magnetization on pyrochlore lattice sites,  
$\langle S_{\textbf{r}\textbf{r}'}^+ \rangle$.
(b) Phase diagram, showing phases listed in Table~\ref{table:parameter}.
At small $J_{zz}$, the formation of a charge--2 condensate 
(dashed grey line) converts the $\text{U(1)}_0$ QSL into  
a $\mathbb{Z}_2$ QSL (``hidden'' phase of Fig.~\ref{fig:schematicdiagram}).
A further charge--1 condensation (dashed red line) separates  
this from the strong--coupling Higgs phase (easy--plane AF).
At larger $J_{zz}$ 
and small $J_
\pm$ there is a 1$^{st}$--order phase transition (solid red line) 
into phases where $\langle Q_\textbf{r} \rangle \ne \langle Q_\textbf{r'} \rangle$, which becomes a 2$^{nd}$--order transition at larger $J_\pm$. (dashed red line)  
These comprise a charged version of the $\mathbb{Z}_2$ QSL, 
and supersolid which is the charged version of the Higgs phase.
All phase boundaries were estimated within gauge mean-field theory 
(gMFT) for $\mathcal{H}_{\text{XZZ+}}$ [Eq.~(\ref{eq:H.extended})] 
with ferromagnetic Ising interactons $J_{zz} > 0$, 
$J_\pm > 0$, $u=0.04J_z$  [cf. Fig.~\ref{fig:u=0.04}], as described in 
the Supplementary material.
}
\end{figure}

This intermediate ``hidden'' regime is expected to be a rich source of other 
novel phases, particularly once higher--order corrections in Eq.~(\ref{eq:H.rotor.extended})
are taken into account.
This is particularly true for frustrated exchange $J_\pm <0$, where the $U(1)$ QSL occurs 
with $\pi$--flux  \citep{Lee2012},  
and the Higgs transition is postponed to much larger values of $J_\pm$, 
allowing for larger fluctuations of charge $Q$ [Fig.~\ref{fig:u=0.04}]. 
In particular, where the pairing of spinons includes an offsite component, 
$\Delta_{\mu-\nu}=\langle \phi_\textbf{r}\phi_{\textbf{r}+\textbf{e}_{\mu}-\textbf{e}_{\nu}}\rangle$,
the $\mathbb{Z}_2$ QSL will develop a quadrupole moment on bonds
$\langle S_i^+S_j^+\rangle \sim \Delta_0^*\Delta_{\mu-\nu}$, leading to a state with spin--nematic 
order \cite{Andreev1984,Chubukov1991,Shannon2006,Shindou2009,Shindou2011,Kohama2019}.
\label{sec:Supersolid phases}
%
{\textbf{\textit { Supersolid phases ---}}} 
So far, we have shown that $\delta\mathcal{H}_{\text{rotor}}$ qualitatively modifies 
the spinon action $\mathcal{H}_{\text{rotor}}\rightarrow \mathcal{H}_{\text{rotor}}+\mathcal{H}_u$,  
leading to the phase diagram Fig.~\ref{fig:schematicdiagram}. 
%
Still more new phases can be anticipated where a charge instability $\langle Q_\textbf{r}\rangle\neq 0$ is encountered. 
With this in mind, we consider an extended model
\begin{eqnarray}
H_{\text{XXZ}+} = H_{\text{XXZ}} + H_{\text{zz}} \;,
\label{eq:H.extended}
\end{eqnarray}
where $H_{\text{XXZ}}$ is defined in Eq.~(\ref{eq:H.XXZ}), 
and $H_{\text{zz}}$ are further--neighbor Ising interactions 
\citep{Rau2016, Udagawa2016}
\begin{eqnarray}
H_{\text{zz}}=-J_{zz}\sum_{\substack{i\in\vartriangle_\textbf{r} \\  j\in\triangledown_{\textbf{r}'}}} S_i^zS_j^z \rightarrow \mathcal{H}_{\text{ZZ}}=J_{zz}\sum_{\langle \textbf{r}\textbf{r}'\rangle}Q_\textbf{r}Q_{\textbf{r}'}, \quad
\label{eq:H.ZZ}
\end{eqnarray}
and the sum on $i,j$ runs over up/down tetrahedra corresponding to sites 
$\vartriangle_\textbf{r}/\triangledown_{\textbf{r}'}$ [Fig. \ref{fig:QQ}]. 
(In terms of the pyrochlore--lattice, this comprises all 2$^{nd}$--neighbor 
bonds and a subset of 3$^{rd}$--neighbor ones).
We consider the further--neighbor Ising interactions to be FM ($J_{zz} > 0$), 
implying that the resulting interactions between charges $Q_\textbf{r}$ are repulsive.


We  use the projected rotor formalism, Eq.~(\ref{eq:S'}), to explore the new 
instabilities which arise in the extended XXZ model, Eq.~(\ref{eq:H.extended}), 
close to the phase boundaries already identified in Fig.~\ref{fig:u=0.04}, 
with results  summarised in Fig.~\ref{fig:ChargeInt}.
The full solution of the original spin model, for more general parameters, 
is left as an open problem.
Within the rotor framework, for $J_{\pm}/J_z \ll 1$ (and fixed $u$), 
the leading effect of $H_{\text{zz}}$ [Eq.~(\ref{eq:H.ZZ})] 
is to renormalise the (Coulomb) interaction between gauge charges, 
reducing the charge stiffness [See supplemental material for details].  
This increases charge fluctuations, and shifting the boundary between 
the $U(1)$ and $\mathbb{Z}_2$ QSL's to smaller values of $J_\pm$, 
and eventually eliminating the $U(1)$ QSL entirely for $J_{zz} /J_z \gtrsim 0.2$.

\begin{table}
\centering
\caption{
Phases found in extended XXZ model 
$H_{\text{XXZ}+}$ [Eq.~(\ref{eq:H.extended})], 
as shown in phase diagram Fig.~\ref{fig:ChargeInt}.
Order parameters are listed, as well as the associated experimental signatures
in heat capacity $c(T)$, and spin structure factor $S(\textbf{q})$.}
\label{table:parameter}
\begin{tabular}{c|ccccc}
\noalign{\smallskip}\noalign{\smallskip}\hline\hline
& $\langle \phi_\textbf{r}\rangle$ & $\langle \phi_\textbf{r}\phi_\textbf{r}\rangle$ & $\delta Q$ & $c(T)$ & $S(\textbf{q})$  \\
\hline
$\text{U(1)}_0$ QSL & 0 & 0 & 0 & Power-law & Diffuse \\
$\mathbb{Z}_2$ QSL & 0 & $\neq 0$ & 0 &   Expo-decay & Diffuse\\
$\text{Higgs}_0$  & $\neq 0$ & $\neq 0$ & 0 &   Power-law & Bragg Peaks \\
Charged $\mathbb{Z}_2$ QSL  & 0 & $\neq 0$ & $\neq 0$ &  Expo-decay & Bragg Peaks\\
Supersolid  & $\neq 0$ & $\neq 0$ & $\neq 0$ &   Power-law & Bragg Peaks\\
\hline
\hline
\end{tabular}
\end{table}


With further increase, for $J_{zz}/J_z \gtrsim 1/4$, $H_{\text{zz}}$ [Eq.~(\ref{eq:H.ZZ})] 
drives an instability against a state with a finite charge polarisation 
$\delta Q \equiv \langle Q_{\textbf{r}\in A} \rangle = -\langle Q_{\textbf{r}\in B} \rangle$, 
which breaks the inversion and time--reversal symmetries of the pyrochlore lattice, 
without breaking translational symmetry.
Close to this phase boundary, the ground state energy can be estimated as
%
%
 \begin{eqnarray}
E[\delta Q]=(J_z-4J_{zz})(\delta Q)^2+ \frac{1}{N}\langle \delta\mathcal{H}_{\text{rotor}}\rangle.
\label{eq:GSzz}
\end{eqnarray}
Within a mean--field theory for the pure rotor model, Eq.~(\ref{eq:H.rotor}), in the static limit 
$\delta Q (\omega_n=0)$, spinon and gauge charge degrees of freedom 
are completely decoupled.
In this case, as in the  the classical limit $J_\pm =0$ \citep{Rau2016,Udagawa2016}, the 
polarisation is either zero ($\delta Q =0$), or takes on its maximum possible value 
($\delta Q \approx \delta Q|_{\text{max}}$).
However introducing an interaction between the gauge charge and spinons, 
$\delta\mathcal{H}_{\text{rotor}}$ [Eq.~(\ref{eq:delta.H.rotor})] changes this, allowing 
for intermediate values $0 < \delta Q < \delta Q|_{\text{max}}$, and permitting 
different forms of order to coexist.
The driving force for this is a gain in the kinetic energy of the spinons \citep{Bojesen2017}.
As a result, the charge polarization is dressed with spinons, which can condense 
as pairs to give a charged version of the $\mathbb{Z}_2$ QSL, or individually, 
to give a state with supersolid character, cf. Fig.~(\ref{fig:ChargeInt}).


These phases could be distinguished in experiment through differences in 
heat capacity, and the fact that the charge polarisation 
$\delta Q$ induces a dipole moment, which could be detected as a Bragg 
peak in polarized neutron scattering \citep{Maleev_2002, Chang_2010}
--- cf. Table.~\ref{table:parameter}.
Meanwhile, inelastic scattering would reveal a continuum of excitations 
associated with the remaining spinon degrees of freedom. 
While this analysis has been developed for a specific form of interaction, 
Eq. (\ref{eq:H.ZZ}), the route outline from a spin liquid to coexisting orders 
is far more general, and is a compelling manifestation of the quantum 
nature of the problem.

\label{sec:Discussion}
{\textbf{\textit {  Discussion ---}}}
In this Letter, we have developed a systematic method 
of unveiling the unusual, ``hidden'' phases which descend 
from QSLs described by a lattice gauge theory.
The mechanism we identify is the effective 
interactions between fractionalised quasi--particles (partons), 
which are generated by the physical constraint on the HiIlbert 
space of the parent spin Hamiltonian.
While earlier theoretical works \citep{Fradkin1990, Moessner2003, Hermele2004} 
typically used pertubation theory to address the domain near to a soluble point, 
this approach makes it possible to connect the confined and deconfined limits
of the lattice gauge theory, and to explore the new phases which arise 
at intermediate coupling.
Generically, we find that these effective interactions can lead to a partial lifting 
of gauge symmetry, prior to the onset of a fully--confined, Higgs phase.
In the specific case of pyrochlore quantum spin ice, this takes the form of a 
$\mathbb{Z}_2$ QSL, intermediate between a $U(1)$ QSL, and an 
easy--plane antiferromagnet, in which gauge fluctuations are fully confined.


We anticipate that this approach will make it possible to explore potential new phases
in models which are difficult to solve by other methods.
These include quantum spin ice models with frustrated interactions, where 
a $U(1)$ QSL with $\pi$-flux is expected at small $|J_\pm|/J_z$ \cite{Lee2012}, but 
quantum Monte Carlo (QMC) simulation fails, leaving the properties 
of this state relatively unexplored.
Here we take encouragement from recent numerical results: 
variational calculations identify both the $\pi$-flux QSL, and 
a quantum spin nematic descended from it, at larger values of $|J_\pm|/J_z$  \citep{Benton2018}.
%
%
Moreover, a $\mathbb{Z}_2$ QSL, of the type we predict, has also been identified 
in QMC simulations of an extended model of unfrustrated quantum spin ice, 
occurring intermediate between a $U(1)$ QSL and its fully--confined Higgs phase  
\citep{Huang2020}.


Research into QSL continues to flourish, as new discoveries follow in both 
experiment and theory.
The results in this Letter suggest that each new spin liquid discovered 
represents not only an opportunity in itself, but also a gateway to 
other new phases, which may have properties as exotic and 
interesting as the QSL they descend from.
The approach developed here is a applicable to a wide range 
of spin liquids, and as such it should provide a valuable guide 
in the search for new quantum phases of matter.

 
\label{sec: Acknowledgments}
{\textbf{\textit {Acknowledgments ---}}}
We would like to thank Leon Balents, Yong Baek Kim, Owen Benton, 
Gang Chen, Han Yan, GiBaik Sim and Hee Seung Kim 
for helpful discussions 
and comments on an early draft of the manuscript.
This work is supported by National Research Foundation Grant 
(NRF-2020R1F1A1073870, NRF-2020R1A4A3079707), 
and by the Theory of Quantum Matter Unit, Okinawa Institute of 
Science and Technology Graduate University.

\bibliography{QSL_hidden-v5}

\setcounter{equation}{0}
\setcounter{figure}{0}
\setcounter{table}{0}
\renewcommand{\theequation}{S\arabic{equation}}
\renewcommand{\thefigure}{S\arabic{figure}}

\pagebreak
\newpage

\thispagestyle{empty}
\mbox{}
\pagebreak
\newpage
\onecolumngrid
\begin{center}
  \textbf{\large Hidden phases born of a quantum spin liquid: Application to pyrochlore spin ice\\Supplementary Information}\\[.2cm]
  
  Hyeok-Jun Yang,$^{1}$ Nic Shannon,$^{2}$ and SungBin Lee$^1$\\[.1cm]
  {\itshape ${}^1$Department of Physics, Korea Advanced Institute of Science and Technology, Daejeon, 34141, Korea\\
  \itshape ${}^2$Theory of Quantum Matter Unit, Okinawa Institute of Science and Technology Graduate University,Onna-son, Okinawa 904-0412, Japan\\
}
(Dated: \today)\\[1cm]
\end{center}
\onecolumngrid

\section{I. Review of U(1) Lattice gauge theory}
\label{sec: Review of U(1) Lattice gauge theory}

Here we briefly review the compact U(1) lattice gauge theory of quantum spin ice, following Refs. \citep{Savary2012, Lee2012, Gingras2014, Savary2016}. 
We concentrate on the low--energy physics of the U(1) QSL, which is central to our analysis. 
We consider a pseudospin-1/2 model 
\begin{eqnarray}
	H_{\text{XXZ}} &=& H_0 + H_1  \; ,
\label{eq.minimal.model} 
\end{eqnarray}
[Eq.~(\ref{eq:H.XXZ}) of main text], consisting of a dominant Ising exchange term 
\begin{eqnarray}
	H_0 &=& J_z \sum_{\langle ij \rangle} S_i^z S_{j}^z 
\label{eq:H0} 
\end{eqnarray}
and a quantum fluctuation term 
\begin{eqnarray}
	H_1 &=& - J_{\pm} \sum_{\langle ij \rangle}   (S_i^+S_{j}^- + \text{h.c.}) \; .
\label{eq:H1} 
\end{eqnarray}
Classical spin ice configurations satisfying a two--in, two--out condition 
\bea
\sum_{\textbf{r}'\in \langle \textbf{r}\textbf{r}'\rangle}S_{\textbf{r}\textbf{r}'}^z=\sum_{\mu=0}^3S_{\textbf{r},\textbf{r}+\textbf{e}_\mu}^z=0 \;, 
\eea
minimize
\bea
H_0=J_z\sum_{\langle ij \rangle} S_i^zS_j^z = \frac{J_z}{2}\sum_\textbf{r}\Big(\sum_{\textbf{r}'\in \langle \textbf{r}\textbf{r}'\rangle}S_{\textbf{r}\textbf{r}'}^z \Big)^2 + \text{const.} \;, \quad
\label{eq:Icerule}
\eea
where $\textbf{r}, \textbf{r}'$ are dual lattice (diamond) sites whose bond center is the pyrochlore site $i$. 
We introduce a rotor representation, 
\begin{eqnarray}
&&Q_{\textbf{r}}=\eta_\textbf{r}\sum_{\mu=0}^3S_{\textbf{r},\textbf{r}+\eta_\textbf{r} \textbf{e}_\mu}^z, \quad S_{\textbf{r}\textbf{r}'}^z=\eta_\textbf{r} E_{\textbf{r}\textbf{r}'} \;,
\nonumber\\
&& S_{\textbf{r}\textbf{r}'}^+ =\phi_\textbf{r}^{\dagger} s_{\textbf{r}\textbf{r}'}^+ \phi_{\textbf{r}'} \;\; (s_{\textbf{r}\textbf{r}'}^+=\frac{1}{2}e^{iA_{\textbf{r}\textbf{r}'}}\;\; \text{for}\; \textbf{r}\in A) \;,
\label{eq:eq2}
\end{eqnarray}
[Eq.~(\ref{eq:OriRotor}) of main text] in which the local constraint is mapped to 
the charge-free condition $Q_\textbf{r}=0$ for all $\textbf{r}$ by the Gauss law. 
Similarly, Eq.~(\ref{eq:eq2}) express the spin operators to be gauge-invariant under the local transformation 
\begin{eqnarray}
\phi_\textbf{r} \rightarrow \phi_\textbf{r} e^{ i\Lambda_\textbf{r}},\quad A_{\textbf{r}\textbf{r}'}\rightarrow A_{\textbf{r}\textbf{r}'}+(\Lambda_\textbf{r}-\Lambda_{\textbf{r}'}) \;.
\end{eqnarray}
From now on, we distinguish the Hamiltonian in mathcal font when it is represented in terms of the rotor variables Eq. (\ref{eq:eq2}) while the original spin Hamiltonian Eqs. (\ref{eq.minimal.model})-(\ref{eq:H1}) is written in capital letter.

The divergenceless condition is violated by the spin flipping term $H_1$, which creates and annihilates a pair of gapped spinon excitations. In Eq. (\ref{eq:eq2}), the spin flip $S_{\textbf{r}\textbf{r}'}^+$ raises (lowers) the gauge charge at the diamond sites $\textbf{r}$ ($\textbf{r}'$) by $\pm 1$, in other words 
\bea
[\phi_\textbf{r},Q_\textbf{r}]=\phi_\textbf{r} \;\text{ or }\; [\varphi_\textbf{r}, Q_\textbf{r}]=i \;,
\label{eq:commu}
\eea
where $\phi_\textbf{r}=e^{-i\varphi_\textbf{r}}$.
The virtual excitation by spin flip term $H_1$ lifts the extensive degeneracy of classical spin ices and gives rise to the electromagnetic energy in the low-energy sector.
\begin{eqnarray}
\mathcal{H}_{\text{EM}}=\frac{U}{2}\sum_{\langle \textbf{r}\textbf{r}' \rangle} E_{\textbf{r}\textbf{r}'}^2 -\sum_{\hexagon}g_p \cos(\mathcal{B}_p) \;,
\label{eq:EM}
\\
\mathcal{B}_p=\sumcirclearrowleft_{\langle \textbf{r}\textbf{r}'\rangle \in \hexagon}A_{\textbf{r}\textbf{r}'}, \quad g_p=3J_{\pm}^3/2J_z^2 \;,  
\label{eq:gp}
\end{eqnarray}
where the first term in Eq. (\ref{eq:EM}) enforces the electric field to be 
\bea
E_{\textbf{r}\textbf{r}'}=\eta_\textbf{r} S_{\textbf{r}\textbf{r}'}^z=\pm 1/2 \quad \text{for large}\;\; U>0 \;,
\eea
and the symbol $\hexagon$ denotes the hexagonal plaquette which the lattice curl of the gauge field $A_{\textbf{r}\textbf{r}'}$ is defined on. 
For the unfrustrated $J_{\pm}>0$, the coupling constant $g_p$ stabilizes 0-flux, $\mathcal{B}_p=0$. Meanwhile the frustrated exchange $J_{\pm}<0$ stabilizes the $\pi$-flux, $\mathcal{B}_p=\pi$, which doubles the unit cell.

Along with the gapped spinon, there are two more excitations in Eq. (\ref{eq:EM}). Since the gauge field $A_{\textbf{r}\textbf{r}'} (\text{mod}\; 2\pi)$ is compact, the magnetic energy allows the topological defect, a gapped magnetic monopole excitation. Also the quantum theory of Eq. (\ref{eq:EM}) quantizes the field variables, 
\bea
[A_{\textbf{r}\textbf{r}'},E_{\textbf{r}\textbf{r}'}]=-i \quad \text{on the same link}\; \textbf{r}\textbf{r}'.
\eea
In the continuum limit, Eq. (\ref{eq:EM}) becomes
\bea
\mathcal{H}_{\text{EM}}\sim E_{\textbf{r}\textbf{r}'}^2+\mathcal{B}_p^2 \;,
\eea
which is analogous to the harmonic oscillator whose quanta corresponds to the gapless photon.

The total partition function, taking into account both partons and gauge fields, 
is then defined as
\bea
\mathcal{Z}_{\text{total}}=\int\mathcal{D}\phi_{\textbf{r}}^*\mathcal{D}\phi_{\textbf{r}}\mathcal{D}Q_{\textbf{r}}\mathcal{D}A_{\textbf{r}\textbf{r}'}\delta(\sum_{\mu=0}^3E_{\textbf{r},\textbf{r}+\eta_{\textbf{r}}e_{\textbf{r}}}-Q_{\textbf{r}})
\exp\Big[
-S_{\text{eff}} [\phi_{\textbf{r}}^*,\phi_{\textbf{r}},Q_{\textbf{r}}]
-S_{\text{EM}} [A_{\textbf{r}\textbf{r}'}]
\Big] \; ,
\label{eq:S.full.model}
\eea
where $S_{\text{EM}}[A_{\textbf{r}\textbf{r}'}]$ is the pure electromagnetic action defined 
in Eq.~(\ref{eq:EM}), and $S_{\text{eff}}$ describes a quantum rotor model 
\bea
S_{\text{eff}} &=& \int_0^{\beta} d\tau \Big(\sum_\textbf{r}(iQ_\textbf{r}\partial_{\tau}\varphi_\textbf{r} +\lambda(\phi_\textbf{r}^*\phi_\textbf{r}-1)) + \mathcal{H}_{\text{rotor}}\Big)  \; ,
\label{eq:S.rotor.model}
\eea
[Eq.~(\ref{eq:S.rotor}) of the main text], where
\bea
\mathcal{H}_{\text{rotor}} &=& \frac{J_z}{2}\sum_\textbf{r}  Q_\textbf{r}^2 
- \frac{J_\pm}{4}\sum_{\langle\langle\textbf{r}\textbf{r}''\rangle\rangle} \phi_{\textbf{r}}^{\dagger}e^{i( A_{\textbf{r}\textbf{r}'}+A_{\textbf{r}'\textbf{r}''})} \phi_{\textbf{r}''} +\text{h.c.} \; , 
\eea
[Eq.~(\ref{eq:H.rotor}) of the main text].
We note that the delta function in Eq.~(\ref{eq:S.full.model}) 
implements the Gauss law associated with the gauge charge.
This restricts fluctuations to the physical Hilbert space $|Q_\textbf{r}|< 4S$, 
as pointed out in the main text.

In the limit of weak gauge field fluctuations, one can fix the specific 
gauge in Eqs.~(\ref{eq:EM}) and (\ref{eq:gp}).
Doing so, while allowing the gauge charge to take on values $Q \in (-\infty, \infty)$, 
we arrive the effective parton action for a pure quantum rotor model, 
\begin{eqnarray}
&&\mathcal{Z}=\int \mathcal{D}\phi_\textbf{r}^*\mathcal{D}\phi_\textbf{r} \mathcal{D}Q_\textbf{r}|_{Q_\textbf{r}\in (-\infty,\infty)} e^{-\mathcal{S}_{\text{eff}}[\phi_\textbf{r}^*,\phi_\textbf{r},Q_\textbf{r}]} \; , 
\end{eqnarray}
as defined in Eq.~(\ref{eq:Z.rotor})--(\ref{eq:H.rotor}) of the main text.

This quantum rotor formalism also encompasses the emergence of magnetic order through 
a Higgs transition.
Within the U(1) QSL which occurs $J_z\gg|J_\pm|$, spinons are both gapped and deconfined.
As $|J_\pm|$ is increased, the spinon bandwidth increases, and the gap to spinon eventually closes, 
signalling the spinon condensation $\langle \phi_\textbf{r}\rangle \neq 0$. 
In the fully--confined limit $|J_\pm|\gg J_z$, the U(1) gauge structure is broken and the 
artificial photon is gapped out by Higgs mechanism. 
The ground state no longer posseses long--range entanglement, and is adiabatically connected to 
a product state with in--plane magnetization,  
\bea
\langle S_{\textbf{r}\textbf{r}'}^+\rangle=\langle \phi_\textbf{r}^{\dagger}\rangle \langle s_{\textbf{r}\textbf{r}'}^+\rangle \langle \phi_{\textbf{r}'}\rangle \neq 0 \;. 
\eea
For the minimal model, Eq.~(\ref{eq.minimal.model}), these ground states are conventional--magnetically ordered states with ordering wave vector $\textbf{k}_0=(0,0,0)$ for unfrustrated exchanges, $J_\pm > 0$, and $\textbf{k}_{\pi}=(k,0,0)$ for frustrated exchange, 
$J_\pm < 0$ \cite{Lee2012}.

Once more general  (anisotropic) exchange models are considered, a more diverse 
set of possible magnetic orders results. 
For example, terms of the form $\sim J_{\pm z}S_i^+S_j^z$ can stabilize the coexistence 
of the deconfined spinons and the Ising orders, while terms of the form $\sim J_{\pm\pm}S_i^+S_j^+$ 
tend to condense the spinons at $\textbf{k}=(2\pi,0,0)$ \cite{Lee2012}.
In this work, however, we concentrate instead on the possibility of phase transitions which are not 
driven by the condensation of a single spinon, and so lead to other, more exotic, phases of matter.

\section{II. Projected rotor representation}
\label{sec:Projected.rotor.representation}

We now discuss the consequences of restricting fluctuation of charge to the physical 
domain $Q_\textbf{r} \in (-4S, 4S)$, within the rotor formalism, deriving the effective interaction
between spinons given in Eq.~(\ref{eq:parameter.u}) of the main text.
The recipe follows the two steps, (i) the restriction of the enlarged Hilbert space to be the physical one $\mathbb{H}_S$ based on Eq. (\ref{eq:OriRotor}), (ii) then the Hamiltonian is modified $\mathcal{H}_{\text{rotor}}\rightarrow \mathcal{H}_{\text{rotor}}'$ to take account of finiteness of the Hilbert space. 

For simplicity, we first restrict the domain $Q_{\textbf{r}}\in (-4S,4S)$ at a site $\textbf{r}$ only and leave others unchanged $Q_{\textbf{r}'\neq \textbf{r}} \in (-\infty,\infty)$. 
Our strategy is to obligate the divergent eigenenergies of $\mathcal{H}_{\text{rotor}}'$ to suppress the Boltzmann factors outside the domain $Q_{\textbf{r}}\in (-4S,4S)$. 
Then we look for the Hermitian operator $P_{\textbf{r}}$ defined by 
\bea
\mathcal{H}_{\text{rotor}}'\equiv P_{\textbf{r}}\mathcal{H}_{\text{rotor}}P_{\textbf{r}} \;,
\eea
ensuring Eq. (\ref{eq:S'}). Since $P_\textbf{r}$ diagnoses whether $|Q_\textbf{r}|<4S$ or not, it would be a function of $|Q_\textbf{r}|$. 
\bea
|P_{\textbf{r}}[Q_{\textbf{r}}]|^2&&\overset{?}{=}\Theta[4S-|Q_{\textbf{r}}|]+ \Theta[-4S+|Q_{\textbf{r}}|] \cdot \infty , 
\nonumber\\
  \Theta[x] &&=
  \begin{cases}
    1 & \text{for } x>0 \\
    0 & \text{otherwise}  \; ,
  \end{cases} 
\label{eq:P1}
\eea
However, the function Eq. (\ref{eq:P1}) is ill-defined due to the ambiguities in the second term and 
at $|Q_{\textbf{r}}|=4S$.
From the mapping Eq. (\ref{eq:OriRotor}), the gauge charge $Q_\textbf{r}$ actually takes 
the integer values $0, \pm 1,... \pm 4S$ only. 
Instead of Eq. (\ref{eq:P1}), we consider the function $P_{\textbf{r}}[Q_{\textbf{r}}]$ of 
to real $Q_{\textbf{r}}\in (-\infty,\infty)$. 
Then we can define
\bea
P_{\textbf{r}}[Q_{\textbf{r}}]=
  \begin{cases}
    1 & \text{for } Q_{\textbf{r}}=0, \pm 1, ... , \pm 4S \\
    0 & \text{for } Q_{\textbf{r}}=\pm (4S+1) \; ,
  \end{cases}
\label{eq:Pcondition}
\eea
The first condition for $Q_\textbf{r}=0,\pm 1,...,\pm 4S$ validates the first term of Eq.~(\ref{eq:P1}). 
The second condition $P_\textbf{r}[Q_\textbf{r}=\pm(4S+1)]=0$ is resolved as
\bea
&&P_\textbf{r}\phi_{\textbf{r}}^\dagger P_\textbf{r}|Q_\textbf{r} \rangle =
  \begin{cases}
    |Q_\textbf{r}+1\rangle & \text{for } Q_{\textbf{r}}=0, \pm 1, ... , \pm (4S-1), -4S \\
    0 & \text{for } Q_{\textbf{r}}=+4S
  \end{cases}
\nonumber\\
&&P_\textbf{r}\phi_{\textbf{r}} P_\textbf{r}|Q_\textbf{r} \rangle =
  \begin{cases}
    |Q_\textbf{r}-1\rangle & \text{for } Q_{\textbf{r}}=0, \pm 1, ... , \pm (4S-1), +4S \\
    0 & \text{for } Q_{\textbf{r}}=-4S
  \end{cases}
\nonumber\\
&&P_\textbf{r}Q_{\textbf{r}} P_\textbf{r}|Q_\textbf{r} \rangle =
  \begin{cases}
    Q_\textbf{r}|Q_\textbf{r}\rangle & \text{for } Q_{\textbf{r}}=0, \pm 1, ... , \pm 4S \\
    0 & \text{for } Q_{\textbf{r}}=\pm (4S+1) \; ,
  \end{cases}
\label{eq:Pladder}
\eea
as implied by the ladder termination
\bea
S_{\textbf{r}\textbf{r}'}^{\pm}|S_{\textbf{r}\textbf{r}'}^z=\pm S \rangle =0 \; .
\label{eq:Sladder}
\eea
To obtain a closed--form expression for the projection operator, we include the  
integers $Q_\textbf{r} = \pm (4S+1)$ in its domain, and seek a polynomial 
satisfying the conditions Eq.~(\ref{eq:Pladder}).
Since Eq. (\ref{eq:P1}) is an even function of $Q_\textbf{r}$, this polynomial 
$P_{\textbf{r}}$ also consists of even powers of $Q_\textbf{r}$.
Defined in this way, there is one and only one solution for $P_{\textbf{r}}$ 
to Eq. (\ref{eq:Pcondition}), a polynomial of order $2(4S+1)$
\bea
P_{\textbf{r}}[Q_{\textbf{r}}]&&= 1-\frac{1}{(4S+1)(8S+1)!}\prod_{k=0}^{4S}(Q_{\textbf{r}}^2-k^2)
\nonumber\\
&&=1-\alpha_S Q_{\textbf{r}}^2+...-\frac{1}{(4S+1)(8S+1)!}Q_{\textbf{r}}^{8S+2} \quad\quad (\alpha_S =\frac{((4S)!)^2}{(4S+1)(8S+1)!}) \; .
\label{eq:P2}
\eea

\begin{figure}[t!]
\subfloat[]{\label{fig:P}\includegraphics[width=0.5\textwidth]{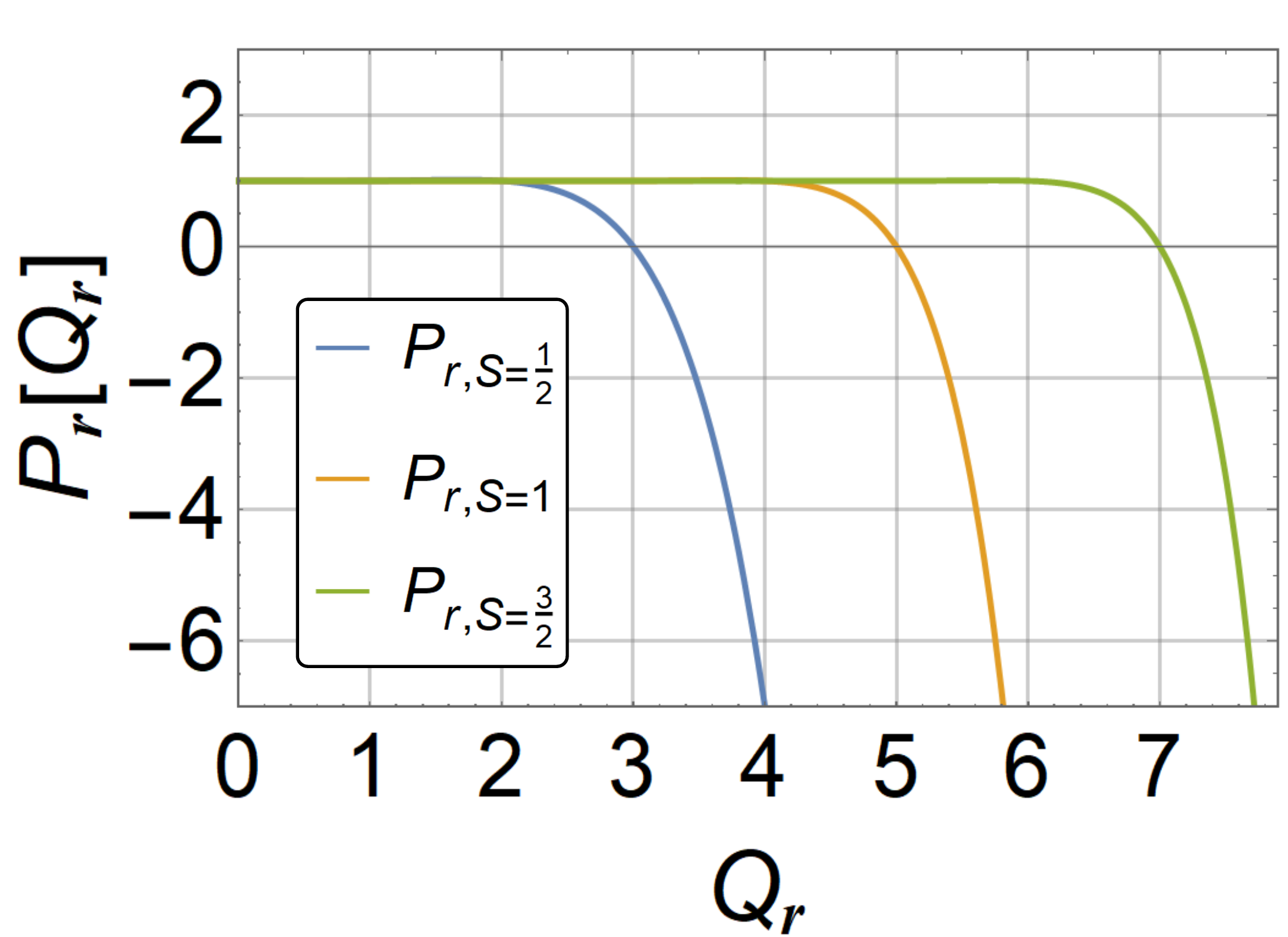}}
\subfloat[]{\label{fig:P^2}\includegraphics[width=0.5\textwidth]{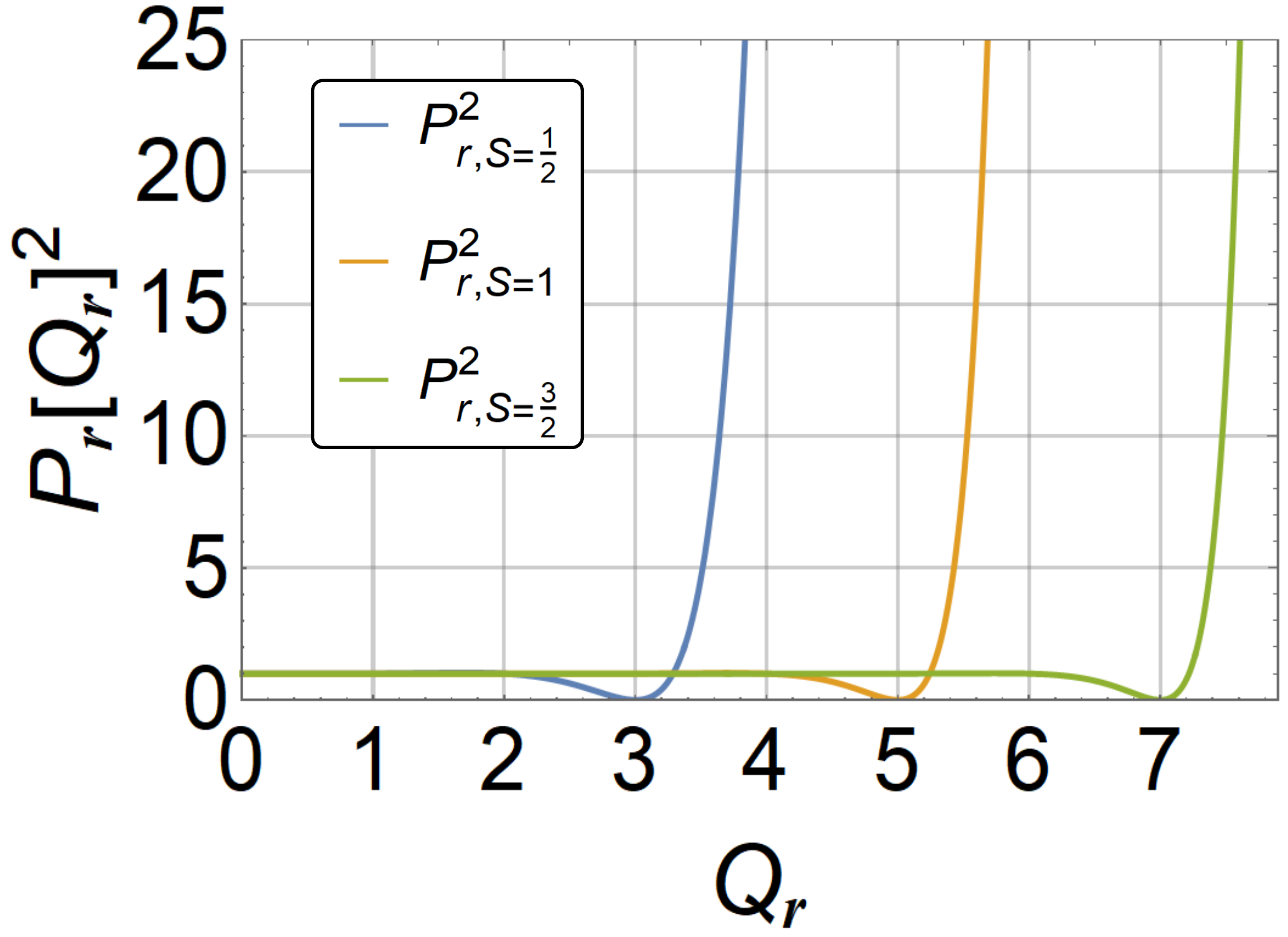}}
\caption{Projector operators $P_{\textbf{r}}[Q_{\textbf{r}}]$ and its square $(P_{\textbf{r}}[Q_{\textbf{r}}])^2$ for the spin $S=1/2, 1, 3/2$ exhibiting (a) the approximated step function inside $|Q_\textbf{r}| \leq 4S$ and (b) divergences outside $|Q_\textbf{r}|>4S$ to expel the Boltzmann factors concerning large $\mathcal{H}_{\text{rotor}}'= P_{\textbf{r}}\mathcal{H}_{\text{rotor}}P_{\textbf{r}}$.}
\label{fig:Projector}
\end{figure}

From this series, we single out the correction with the lowest power, $\alpha_S Q_r^2$.
This is a rapidly decreasing function of the control parameter $S$: by application of the Stirling approximation
$\log(N!) \approx N\log N$ we find
\bea
\alpha_{S\gg 1} {\approx}\frac{1}{S 2^{8S+2}} \; .
\eea
In the physical limit, $S=\frac{1}{2}$, it has the value 
\bea
\alpha_{S=\frac{1}{2}}=\frac{1}{90} \; .
\eea
Most significant is the sign of this term, $\alpha_S >0$.   
This is fixed by the condition Eq.~(\ref{eq:Pcondition}). 
%
Adding additional constraints of the form $P[|Q_{\textbf{r}}| \geq 4S+2]= 0$, for integer $Q_{\textbf{r}}$ outside
the physical range introduce higher--order terms $O(Q_{\textbf{r}}^{8S+2})$ in Eq.~(\ref{eq:P2}), but leaves 
the sign of $\alpha_S$ unchanged.

In Fig.~\ref{fig:P} we plot the polynomial form of the projection operator 
$P_\textbf{r}[Q_\textbf{r}]$ [Eq.~(\ref{eq:P2})] as a function of $Q_\textbf{r}$, for $S=1/2, 1, 3/2$.
Meanwhile, in Fig.~\ref{fig:P^2} we show its square $P_\textbf{r}[Q_\textbf{r}]^2$.
It is the square of the projection operator which determines the matrix elements entering into the 
projected rotor Hamiltonian [Eq.~(\ref{eq:H.rotor.projected}) of the main text].
We see that $P_\textbf{r}[Q_\textbf{r}]^2 \approx 1$ on the physical interval $(-4S, 4S)$, 
but diverges rapidly for $|Q_\textbf{r}| > 4S + 1$, with the asymptotic behaviour 
$P_\textbf{r}[Q_\textbf{r}]^2 \sim Q_\textbf{r}^{16S+4}$ for $|Q_\textbf{r}|\rightarrow \infty$.
This, combined with the $Q_\textbf{r}^2$ term in the original rotor Hamiltonian [Eq.~(\ref{eq:H.rotor}) 
of the main text], will eliminate unphysical states at large $|Q_\textbf{r}|$ from the 
path integral determining the partition function [Eq.~(\ref{eq:S'}) of the main text].


We now generalize the mapping $(Q_\textbf{r}, \phi_\textbf{r}, \phi_\textbf{r}^{\dagger}) \rightarrow (\tilde{Q}_\textbf{r}, \tilde{\phi_\textbf{r}}, \tilde{\phi}_\textbf{r}^{\dagger})$ for all $\textbf{r}$,
\bea
\mathcal{H}_{\text{rotor}}'&&=\mathcal{H}_{\text{rotor}}[\tilde{Q}_\textbf{r},\tilde{\phi}_\textbf{r},\tilde{\phi}_\textbf{r}^{\dagger}] \quad(\tilde{O}_\textbf{r}\equiv P_\textbf{r} O_\textbf{r} P_\textbf{r}), \quad
\nonumber\\
&&=\mathcal{H}_{\text{rotor}}[Q_\textbf{r},\phi_\textbf{r},\phi_\textbf{r}^{\dagger}]+\delta \mathcal{H}_{\text{rotor}}[Q_\textbf{r},\phi_\textbf{r},\phi_\textbf{r}^{\dagger}] \; ,
\label{eq:H'}
\eea
where the new matter fields $\tilde{Q}_\textbf{r}, \tilde{\phi}_\textbf{r}$ project out the unphysical 
portions outside the domain $Q_\textbf{r} \in (-4S, 4S)$. 
This projection generates new effective interactions between the spinons (partons) 
of quantum spin ice.

In Eq. (\ref{eq:H'}), we ignore the non-local interactions coming from $P_{\textbf{r}}\phi_{\textbf{r}'\neq\textbf{r}}^{(\dagger)}P_{\textbf{r}}$ since $\phi_{\textbf{r}'}^{(\dagger)}$ doesn't affect the gauge charge $Q_{\textbf{r}}$ at $\textbf{r}$.
As quantum fluctuation $\mathcal{H}_1$ rises, it monotonically increases $Q_{\textbf{r}}^2$, implying that the corrections $P_\textbf{r}=1-\alpha_S Q_\textbf{r}^2+...$ in Eq. (\ref{eq:P2}) contribute to $\delta H_{\text{rotor}}$ in order of degrees. 
The coefficient $\alpha_S$ in the lowest correction properly recognizes that the effective 
interaction $\delta H_{\text{rotor}}$ is significant for the small spin $S$. 

Integration of $\delta H_{\text{rotor}}$ with respect to the charge 
field $Q_{\textbf{r}}$ contributes to an effective spinon interaction 
in two different ways
\bea 
&&\text{(1) Non-zero commutator Eqs. (\ref{eq:commu}), (\ref{eq:commu2}) during the normal-ordering.}
\nonumber\\
&&\text{(2) Non-trivial Jacobian transformation in the path integral.}
\label{eq:Contributions}
\eea
It turns out that dominant contribution comes from (2), and that the lowest order, 
the effective parton interaction $\delta H_{\text{rotor}}$ is given by
\bea
S_{\textbf{r}\textbf{r}'}^+S_{\textbf{r}'\textbf{r}''}^-&&
\approx \frac{1}{4}\Big(1-\alpha_S(Q_\textbf{r}^2+Q_{\textbf{r}'}^2) \Big)\phi_\textbf{r}^{\dagger}e^{iA_{\textbf{r}\textbf{r}'}}\phi_{\textbf{r}'}
\Big(1-\alpha_S(Q_\textbf{r}^2+2Q_{\textbf{r}'}^2+Q_{\textbf{r}''}^2) \Big)\phi_{\textbf{r}'}^{\dagger}e^{iA_{\textbf{r}'\textbf{r}''}}\phi_{\textbf{r}''}
\Big(1-\alpha_S(Q_{\textbf{r}'}^2+Q_{\textbf{r}''}^2) \Big).
\label{eq:deltaH}
\eea
Physically, the effective parton interaction in Eq. (\ref{eq:deltaH}) reflects that both gauge charges $Q_{\textbf{r}'}=\pm 4S$ block the propagation $\sim \phi_\textbf{r}^{\dagger}\phi_{\textbf{r}'}\phi_{\textbf{r}'}^{\dagger}\phi_{\textbf{r}''}$. Thus it is natural to consider the parton interaction affected by the size of the gauge charge $|Q_{\textbf{r}}|$  (or $Q_{\textbf{r}}^2$) rather than the gauge charge $Q_{\textbf{r}}$ itself.
To evaluate Eq. (\ref{eq:deltaH}) in the path integral formalism, it is required to normally order them. Applying Eq. (\ref{eq:commu}) $n$-times, we obtain
\bea
\phi_r Q_{\textbf{r}}^n=(Q_{\textbf{r}}+1)^n\phi_r, \quad \phi_r^{\dagger}Q_{\textbf{r}}^n=(Q_{\textbf{r}}-1)^n \phi_r^{\dagger} \; .
\label{eq:commu2}
\eea
Before proceeding, we note that the new mapping $S_{\textbf{r}\textbf{r}'}^+=\frac{1}{2}\phi_\textbf{r}^{\dagger}e^{iA_{\textbf{r}\textbf{r}'}}\phi_{\textbf{r}'}\rightarrow\frac{1}{2}\tilde{\phi}_\textbf{r}^{\dagger}e^{A_{\textbf{r}\textbf{r}'}}\tilde{\phi}_{\textbf{r}'}$ now not only raises/lowers the gauge charge at $\textbf{r}/\textbf{r}'$ but also diagnoses whether the gauge charge is within the physical domain $(-4S,4S)$ or not. Thus, the spin exchange Eq. (\ref{eq:deltaH}) depends on the sequences of spin operators $S_{\textbf{r}\textbf{r}'}^+S_{\textbf{r}'\textbf{r}''}^-\neq S_{\textbf{r}'\textbf{r}''}^-S_{\textbf{r}\textbf{r}'}^+$. To address this problem, we evaluate the both contribution to replace $S_{\textbf{r}\textbf{r}'}^+S_{\textbf{r}'\textbf{r}''}^-$ in Eq. (\ref{eq:deltaH}) by the symmetric summation, 
\bea
S_{\textbf{r}\textbf{r}'}^+S_{\textbf{r}'\textbf{r}''}^- \rightarrow \frac{1}{2}(S_{\textbf{r}\textbf{r}'}^+S_{\textbf{r}'\textbf{r}''}^-+S_{\textbf{r}'\textbf{r}''}^-S_{\textbf{r}\textbf{r}'}^+)\;.
\label{eq:symsum}
\eea
As a result, this leads to, 
\bea
\mathcal{H}_1+\delta \mathcal{H}_{\text{rotor}} 
=&& -\frac{J_\pm}{4}\sum_{\langle\langle \textbf{r}\textbf{r}''\rangle\rangle}
\frac{1}{2}\Big(P_{\textbf{r}}[Q_{\textbf{r}}]P_{\textbf{r}}[Q_{\textbf{r}}-1]P_{\textbf{r}'}[Q_{\textbf{r}'}]\underline{(P_{\textbf{r}'}[Q_{\textbf{r}'}+1])^2}P_{\textbf{r}'}[Q_{\textbf{r}'}] P_{\textbf{r}''}[Q_{\textbf{r}''}]P_{\textbf{r}''}[Q_{\textbf{r}''}+1]
\nonumber\\
&&\quad\quad\quad\quad\quad\;\; +P_{\textbf{r}}[Q_{\textbf{r}}]P_{\textbf{r}}[Q_{\textbf{r}}-1]P_{\textbf{r}'}[Q_{\textbf{r}'}]\underline{(P_{\textbf{r}'}[Q_{\textbf{r}'}-1])^2}P_{\textbf{r}'}[Q_{\textbf{r}'}] P_{\textbf{r}''}[Q_{\textbf{r}''}]P_{\textbf{r}''}[Q_{\textbf{r}''}+1] \Big)
\phi_\textbf{r}^{\dagger}e^{i(A_{\textbf{r}\textbf{r}'}+A_{\textbf{r}'\textbf{r}''})}\phi_{\textbf{r}''}+ \text{h.c.}
\nonumber\\ 
= && -\frac{J_\pm}{4}\sum_{\langle\langle \textbf{r}\textbf{r}''\rangle\rangle}
\Big[1-\alpha_S \Big( Q_{\textbf{r}}^2 +2 Q_{\textbf{r}'}^2+Q_{\textbf{r}''}^2 +(Q_{\textbf{r}}-1)^2+(Q_{\textbf{r}'}-1)^2+(Q_{\textbf{r}'}+1)^2 + (Q_{\textbf{r}''}+1)^2
\Big)\Big]
\nonumber\\
&&\times \phi_\textbf{r}^{\dagger}e^{i(A_{\textbf{r}\textbf{r}'}+A_{\textbf{r}'\textbf{r}''})}\phi_{\textbf{r}''}+ \text{h.c.} + O(Q_{\textbf{r}}^4) \;.
\label{eq:deltaHXXZ}
\eea
On the first and the second lines, we mark the underlines to emphasize the different corrections from Eq. (\ref{eq:symsum}).
Integrating out the gauge charge $Q_\textbf{r}$, Eq. (\ref{eq:deltaHXXZ}) gives rise to several powers of $J_\pm \phi_\textbf{r}^{\dagger}e^{i(A_{\textbf{r}\textbf{r}'}+A_{\textbf{r}'\textbf{r}''})}\phi_{\textbf{r}''}$ in the spinon action. Due to the commutator Eq. (\ref{eq:commu2}), the spinon band width is trivially rescaled as 
\bea
J_\pm \rightarrow J_{\pm}(1-4\alpha_S) \;,
\label{eq:rescale}
\eea
in Eq. (\ref{eq:deltaHXXZ}).
The lowest correction signalling non-trivial instabilities out of U(1) QSL comes from the quartic corrections in spinons. Performing the Gaussian integral $\mathcal{H}_0+\delta\mathcal{H}_{\text{rotor}}$ over the charges $Q_\textbf{r}$ and $Q_{\textbf{r}''}$,
\bea
&&\Big(\frac{J_z}{2}+\frac{J_\pm}{4}\mathcal{H}_\textbf{r}\Big)Q_\textbf{r}^2 +Q_\textbf{r}\Big(i\partial_\tau \varphi_{\textbf{r}} -\frac{J_\pm \alpha_S}{2}(\phi_\textbf{r}^{\dagger}e^{i(A_{\textbf{r}\textbf{r}'}+A_{\textbf{r}'\textbf{r}''})}\phi_{\textbf{r}''}-\phi_\textbf{r} e^{-i(A_{\textbf{r}\textbf{r}'}+A_{\textbf{r}'\textbf{r}''})}\phi_{\textbf{r}''}^\dagger)\Big)
\nonumber\\
+&&
\Big(\frac{J_z}{2}+\frac{J_\pm}{4}\mathcal{H}_{\textbf{r}''}\Big)Q_{\textbf{r}''}^2 +Q_{\textbf{r}''}\Big(i\partial_\tau \varphi_{\textbf{r}''} +\frac{J_\pm \alpha_S}{2}(\phi_\textbf{r}^{\dagger}e^{i(A_{\textbf{r}\textbf{r}'}+A_{\textbf{r}'\textbf{r}''})}\phi_{\textbf{r}''}-\phi_\textbf{r}e^{-i(A_{\textbf{r}\textbf{r}'}+A_{\textbf{r}'\textbf{r}''})}\phi_{\textbf{r}''}^\dagger)\Big)
\nonumber\\
\rightarrow \quad && \frac{-1}{2J_z+J_\pm \mathcal{H}_{\textbf{r}}} \Big(i\partial_\tau \varphi_{\textbf{r}} -\frac{J_\pm \alpha_S}{2}(\phi_\textbf{r}^{\dagger}e^{i(A_{\textbf{r}\textbf{r}'}+A_{\textbf{r}'\textbf{r}''})}\phi_{\textbf{r}''}-\phi_\textbf{r}e^{-i(A_{\textbf{r}\textbf{r}'}+A_{\textbf{r}'\textbf{r}''})}\phi_{\textbf{r}''}^\dagger)\Big)^2 
\nonumber\\
&&+ \frac{-1}{2J_z+J_\pm \mathcal{H}_{\textbf{r}''}} \Big(i\partial_\tau \varphi_{\textbf{r}''} +\frac{J_\pm \alpha_S}{2}(\phi_\textbf{r}^{\dagger}e^{i(A_{\textbf{r}\textbf{r}'}+A_{\textbf{r}'\textbf{r}''})}\phi_{\textbf{r}''}-\phi_\textbf{r}e^{-i(A_{\textbf{r}\textbf{r}'}+A_{\textbf{r}'\textbf{r}''})}\phi_{\textbf{r}''}^\dagger)\Big)^2 \;,
\label{eq:GaussianInt}
\eea
where the opposite signs in front of $\sim\phi_\textbf{r}^\dagger\phi_{\textbf{r}''}$ and $\sim\phi_\textbf{r}\phi_{\textbf{r}''}^\dagger$ are resulted from the normal-ordered path integral.
Here, $\mathcal{H}_\textbf{r}[\phi_\textbf{r},\phi_\textbf{r}^{\dagger}]$, the (dimensionless) spinon fields as the coefficient of $Q_\textbf{r}^2$ in Eq. (\ref{eq:deltaHXXZ}) apart from $J_\pm/4$, is defined as folloiwng. 
\bea
&&\mathcal{H}_{\textbf{r}}[\phi_\textbf{r},\phi_\textbf{r}^{\dagger}]=4\alpha_S \Big[ \frac{1}{2}\sum_{\textbf{r}''\in \langle \langle \textbf{r}\textbf{r}''\rangle\rangle}(\phi_\textbf{r}^{\dagger}e^{i(A_{\textbf{r}\textbf{r}'}+A_{\textbf{r}'\textbf{r}''})}\phi_{\textbf{r}''}+ \text{h.c.} ) 
+ \sum_{\substack{\langle \langle \textbf{r}_1 \textbf{r}_2\rangle\rangle 
\\ \textbf{r}_1 \in \langle \textbf{r}\textbf{r}_1\rangle , \;  \textbf{r}_2 \in \langle \textbf{r}\textbf{r}_2\rangle}} (\phi_{\textbf{r}_1}^{\dagger}e^{i(A_{\textbf{r}_1\textbf{r}}+A_{\textbf{r}\textbf{r}_2})}\phi_{\textbf{r}_2}+ \text{h.c.} )
\Big] \;,
\nonumber\\
&&\mathcal{H}_1=-\frac{J_\pm}{4}\sum_{\textbf{r}}\frac{\mathcal{H_{\textbf{r}}}}{8\alpha_S}=-\frac{J_\pm}{4}\sum_{\langle \langle \textbf{r}\textbf{r}''\rangle\rangle}(\phi_\textbf{r}^{\dagger}e^{i(A_{\textbf{r}\textbf{r}'}+A_{\textbf{r}'\textbf{r}''})}\phi_{\textbf{r}''}+ \text{h.c.} ) \;. 
\label{eq:Cr}
\eea
Up to the quartic terms in the spinon fields, Eq. (\ref{eq:GaussianInt}) reads $\sim -(\phi_\textbf{r}^{\dagger}e^{i(A_{\textbf{r}\textbf{r}'}+A_{\textbf{r}'\textbf{r}''})}\phi_{\textbf{r}''})(\partial_\tau \phi_\textbf{r})^2$  and $\sim -(\phi_\textbf{r}^{\dagger}e^{i(A_{\textbf{r}\textbf{r}'}+A_{\textbf{r}'\textbf{r}''})}\phi_{\textbf{r}''})^2$. 
The first of these terms originates in the Berry phase.
For purposes of a mean field theory, in the static limit, this term can be neglected [cf. main text].
The second term, meanwhile, 
originates in the non-zero commutator for the normal-ordered path integral 
in Eqs. (\ref{eq:commu2}) and (\ref{eq:deltaHXXZ}). 
Approximating the denominator in Eq. (\ref{eq:GaussianInt}) as $2J_z$, it becomes,
\bea
\frac{1}{2J_z}\Big( &&(\partial_\tau \varphi_{\textbf{r}})^2 + (\partial_\tau \varphi_{\textbf{r}''})^2-(J_\pm \alpha_S)^2((\phi_\textbf{r}^{\dagger}e^{i(A_{\textbf{r}\textbf{r}'}+A_{\textbf{r}'\textbf{r}''})}\phi_{\textbf{r}''})^2 + (\phi_\textbf{r}e^{-i(A_{\textbf{r}\textbf{r}'}+A_{\textbf{r}'\textbf{r}''})}\phi_{\textbf{r}''}^\dagger)^2 )
\nonumber\\
&&-iJ_\pm \alpha_S(\partial_\tau \varphi_{\textbf{r}} - \partial_\tau \varphi_{\textbf{r}''})(\phi_\textbf{r}^{\dagger}e^{i(A_{\textbf{r}\textbf{r}'}+A_{\textbf{r}'\textbf{r}''})}\phi_{\textbf{r}''}-\phi_\textbf{r}e^{-i(A_{\textbf{r}\textbf{r}'}+A_{\textbf{r}'\textbf{r}''})}\phi_{\textbf{r}''}^\dagger)\Big).
\label{eq:GaussianInt2}
\eea
Where, since $\phi_{\textbf{r}}=e^{-i\varphi_{\textbf{r}}}$, we have used
\bea
&&(\partial_\tau \varphi_{\textbf{r}})^2=\partial_\tau \phi_{\textbf{r}}^*\partial_\tau \phi_{\textbf{r}},
\nonumber\\
&& i(\partial_\tau \varphi_{\textbf{r}} - \partial_\tau \varphi_{\textbf{r}''})(\phi_\textbf{r}^{\dagger}e^{i(A_{\textbf{r}\textbf{r}'}+A_{\textbf{r}'\textbf{r}''})}\phi_{\textbf{r}''}-\phi_\textbf{r}e^{-i(A_{\textbf{r}\textbf{r}'}+A_{\textbf{r}'\textbf{r}''})}\phi_{\textbf{r}''}^\dagger) = \partial_\tau (\phi_\textbf{r}^{\dagger}e^{i(A_{\textbf{r}\textbf{r}'}+A_{\textbf{r}'\textbf{r}''})}\phi_{\textbf{r}''}-\phi_\textbf{r}e^{-i(A_{\textbf{r}\textbf{r}'}+A_{\textbf{r}'\textbf{r}''})}\phi_{\textbf{r}''}^\dagger)\rightarrow 0 \;, \quad\quad
\eea
after the imaginary integration $\int_{0}^{\beta}d\tau  \partial_\tau (\phi_\textbf{r}^{\dagger}e^{i(A_{\textbf{r}\textbf{r}'}+A_{\textbf{r}'\textbf{r}''})}\phi_{\textbf{r}''})=0$.
Thus, $\mathcal{H}_u$ originates entirely from the third term in Eq.~(\ref{eq:GaussianInt2}). 
Finally, we consider the Jacobian contribution from the case (2) in Eq. (\ref{eq:Contributions}). While performing the Gaussian integration in Eq. (\ref{eq:GaussianInt}),
\bea
&&\int \mathcal{D}\phi_\textbf{r}^*\mathcal{D}\phi_\textbf{r} \mathcal{D}Q_\textbf{r} \exp \Big[-\int_0^\beta d\tau \Big\lbrace\sum_\textbf{r}(\frac{J_z}{2}+\frac{J_\pm}{4}\mathcal{H}_\textbf{r}) Q_\textbf{r}^2+iQ_{\textbf{r}}\partial_\tau \varphi_{\textbf{r}}  + ... \Big\rbrace\Big] \;,
\nonumber\\
=&&\int \mathcal{D}\phi_\textbf{r}^*\mathcal{D}\phi_\textbf{r} \mathcal{D}\Big(\frac{\bar{Q}_\textbf{r}}{(1+\frac{J_\pm}{2J_z}\mathcal{H}_{\textbf{r}})^{1/2}} \Big)\exp \Big [ -\int_0^\beta  d\tau \Big\lbrace\sum_\textbf{r}\bar{Q}_\textbf{r}^2+i\frac{\bar{Q}_{\textbf{r}}}{(\frac{J_z}{2}+\frac{J_\pm}{4}\mathcal{H}_{\textbf{r}})^{1/2}}\partial_\tau \varphi_{\textbf{r}}  + ... \Big\rbrace   \Big] \;,
\eea
with the variable transformation $\bar{Q}_\textbf{r}=Q_{\textbf{r}}\sqrt{\frac{J_z}{2}+\frac{J_\pm}{4}\mathcal{H}_{\textbf{r}}}$. 
While the integrand in the exponential are already taken into account in Eqs. (\ref{eq:GaussianInt})-(\ref{eq:GaussianInt2}),
the Jacobian of spinon fields also gives rise to non-trivial contributions to the spinon action. The variable transformation is given by,
\bea
\bar{\phi}_\textbf{r}=\frac{\phi_\textbf{r}}{(1+\frac{J_\pm}{2J_z}\mathcal{H}_\textbf{r}[\phi_\textbf{r},\phi_\textbf{r}^*])^{1/4}}, \quad \bar{\phi}_\textbf{r}^*=\frac{\phi_\textbf{r}^*}{(1+\frac{J_\pm}{2J_z}\mathcal{H}_\textbf{r}[\phi_\textbf{r},\phi_\textbf{r}^*])^{1/4}} \;. 
\label{eq:spinonbar}
\eea
Thus, the inverse transformation,
\bea
&&\phi_\textbf{r} \simeq \frac{\bar{\phi}_\textbf{r}}{(1-\frac{J_\pm}{2J_z}\mathcal{H}_\textbf{r}[\bar{\phi}_\textbf{r},\bar{\phi}_\textbf{r}^*])^{1/4}}\approx \bar{\phi}_\textbf{r}(1+\frac{J_\pm}{8J_z}\mathcal{H}_\textbf{r}[\bar{\phi}_\textbf{r},\bar{\phi}_\textbf{r}^*]) \;,
\nonumber\\
&&\phi_\textbf{r}^*\simeq\frac{\bar{\phi}_\textbf{r}^*}{(1-\frac{J_\pm}{2J_z}\mathcal{H}_\textbf{r}[\bar{\phi}_\textbf{r},\bar{\phi}_\textbf{r}^*])^{1/4}}\approx \bar{\phi}_\textbf{r}^*(1+\frac{J_\pm}{8J_z}\mathcal{H}_\textbf{r}[\bar{\phi}_\textbf{r},\bar{\phi}_\textbf{r}^*]) \;,
\label{eq:spinonbarInv}
\eea
enable us to rewrite the partition function in terms of Eq. (\ref{eq:spinonbar}). 
Assuming the correlation of $\phi_\textbf{r}$ is similar to that of $\bar{\phi}_\textbf{r}$ 
in the intermediate coupling regime, the quartic spinon interaction 
$\sim  (J_\pm^2/\alpha_S J_z) \mathcal{H}_\textbf{r}\cdot\mathcal{H}_\textbf{r}$ is determined by,
\bea
\mathcal{Z}'= \int \mathcal{D}\phi_\textbf{r}^*\mathcal{D}\phi_\textbf{r} \mathcal{D}\Big(\frac{\tilde{Q}_\textbf{r}}{(1-\frac{J_\pm}{2J_z}\mathcal{H}_{\textbf{r}})^{1/2}} \Big) e^{-S_0[\phi_\textbf{r}^*,\phi_\textbf{r},\bar{Q}_\textbf{r}]}=\int \mathcal{D}\bar{\phi}_\textbf{r}^*\mathcal{D}\bar{\phi}_\textbf{r}e^{-S[\bar{\phi}_\textbf{r}^*,\bar{\phi}_\textbf{r}]}\approx \int \mathcal{D}{\phi}_\textbf{r}^*\mathcal{D}{\phi}_\textbf{r}e^{-S[{\phi}_\textbf{r}^*,{\phi}_\textbf{r}]} \;,
\label{eq:Jacobian}
\eea
where $S$ can be approximated by substituting Eq. (\ref{eq:spinonbarInv}). 
Doing so, we obtain, 
\bea
&&\mathcal{H}_1=-\frac{J_\pm}{4}\sum_{\langle \langle \textbf{r}\textbf{r}''\rangle\rangle}(\phi_\textbf{r}^*e^{i(A_{\textbf{r}\textbf{r}'}+A_{\textbf{r}'\textbf{r}''})}\phi_{\textbf{r}''}+ \text{h.c.} )
\nonumber\\
\rightarrow \quad &&\mathcal{H}_1+\mathcal{H}_u=-\frac{J_\pm}{4}\sum_{\langle \langle \textbf{r}\textbf{r}''\rangle\rangle}(\bar{\phi}_\textbf{r}^*e^{i(A_{\textbf{r}\textbf{r}'}+A_{\textbf{r}'\textbf{r}''})}\bar{\phi}_{\textbf{r}''}+ \text{h.c.} )-\frac{3J_\pm^2 \alpha_S}{8J_z} \sum_{\langle \langle \textbf{r}\textbf{r}''\rangle\rangle}\Big((\bar{\phi}_\textbf{r}^*e^{i(A_{\textbf{r}\textbf{r}'}+A_{\textbf{r}'\textbf{r}''})}\bar{\phi}_{\textbf{r}''})^2+ \text{h.c.} \Big) +\cdots \text{ } .
\label{eq:Jacobian.con}
\eea

Combining Eqs. (\ref{eq:GaussianInt2}) and (\ref{eq:Jacobian.con}), we find that the rotor model, 
Eq. (\ref{eq:H.rotor}), is modified as follows
\bea
&&S[\phi_\textbf{r},\phi_\textbf{r}^*]=\int_0^{\beta}d\tau \Big[\sum_\textbf{r} \Big(\frac{1}{2J_z}\partial_{\tau}\phi_\textbf{r}^*\partial_{\tau}\phi_\textbf{r} + \lambda(\phi_\textbf{r}^*\phi_\textbf{r} -1)\Big)
 + \mathcal{H}_1+\mathcal{H}_u \Big] \;,
\label{eq:Sspinon}
\\
&&\mathcal{H}_u
=-u_0\sum_{\langle\langle \textbf{r}\textbf{r}''\rangle\rangle}\Big(\phi_\textbf{r}^{\dagger}
e^{i(A_{\textbf{r}\textbf{r}'}+A_{\textbf{r}'\textbf{r}''})}\phi_{\textbf{r}''}\Big)^2 
- u_1
\sum_{\substack{\textbf{r}_1 \neq \textbf{r}_2\\ \textbf{r}_1' = \textbf{r}_2'}}
\Big(\sum_{\substack{\langle\langle \textbf{r}_1\textbf{r}_1''\rangle\rangle}}  
\phi_{\textbf{r}_1}^{\dagger} 
e^{i(A_{\textbf{r}_1\textbf{r}'_1}+A_{\textbf{r}'_1\textbf{r}''_1})}\phi_{\textbf{r}''_1}\Big) 
\Big(\sum_{\substack{\langle\langle \textbf{r}_2\textbf{r}_2''\rangle\rangle}}  
\phi_{\textbf{r}_2}^{\dagger}
e^{i(A_{\textbf{r}_2\textbf{r}'_2}+A_{\textbf{r}'_2\textbf{r}''_2})}\phi_{\textbf{r}''_2} + \text{h.c.} \Big)
\nonumber\\ 
&&\quad\quad\;\;\; -u_2
\sum_{\substack{\textbf{r}, \textbf{r}_1'' \neq \textbf{r}_2''}}
\Big(\sum_{\substack{\langle\langle \textbf{r}\textbf{r}_1''\rangle\rangle}}  \phi_{\textbf{r}}^{\dagger}
e^{i(A_{\textbf{r}\textbf{r}'_1}+A_{\textbf{r}'_1\textbf{r}''_1})}\phi_{\textbf{r}''_1}\Big) 
\Big(\sum_{\substack{\langle\langle \textbf{r}\textbf{r}_2''\rangle\rangle}}  \phi_{\textbf{r}}^{\dagger}
e^{i(A_{\textbf{r}\textbf{r}'_2}+A_{\textbf{r}'_2\textbf{r}''_2})}\phi_{\textbf{r}''_2}+\text{h.c.} \Big)
\nonumber\\
&&\quad\quad\;\;\; -u_3
\sum_{\substack{\textbf{r}_1' =\textbf{r}_2}}
\Big(\sum_{\substack{\langle\langle \textbf{r}_1\textbf{r}_1''\rangle\rangle}}  
\phi_{\textbf{r}_1}^{\dagger} 
e^{i(A_{\textbf{r}_1\textbf{r}'_1}+A_{\textbf{r}'_1\textbf{r}''_1})}\phi_{\textbf{r}''_1}\Big) 
\Big(\sum_{\substack{\langle\langle \textbf{r}_2\textbf{r}_2''\rangle\rangle}}  
\phi_{\textbf{r}_2}^{\dagger}
e^{i(A_{\textbf{r}_2\textbf{r}'_2}+A_{\textbf{r}'_2\textbf{r}''_2})}\phi_{\textbf{r}''_2} + \text{h.c.} \Big)
+\text{h.c.} \;,
\label{eq:Husupp}
\eea
where the coefficients 
\bea
&&u_0=\frac{J_\pm^2}{8J_z}\alpha_S(3+4\alpha_S)
,\quad\quad u_1=\frac{J_\pm^2\alpha_S}{2J_z},\quad\quad   u_2=\frac{J_\pm^2\alpha_S}{8J_z},\quad\quad  u_3=\frac{J_\pm^2\alpha_S}{4J_z} \;,
\eea
are all of order $(\sim J_\pm^2/J_z)$, and can be found by enumerating terms in Eq.~(\ref{eq:Cr}).
For small $|J_\pm|/J_z$, the effective interaction $\mathcal{H}_u$ is irrelevant.
However, at finite $|J_\pm|/J_z$, quantum fluctuations can be substantial, and have 
the potential to qualitatively change the nature of the U(1) QSL, by pairing spinons. 
For simplicity, the discussion of this point in the main text is restricted to the 
on--site term $\mathcal{H}_u$.
However the presence of longer--range attracitve interactions in Eq.~(\ref{eq:Husupp}) is 
not expected to suppress spinon pairing; instead it will contribute to the multipolar character 
of the resulting condensate, discussed below.

\section{III. Pure charge-2 propagation}
\label{sec: Pure charge-2 propagation}

\subsection{Multipolar moment}

Here we explore the connection between spinon pairing and phases 
which exhibit a finite multipole moment on bonds. 
%
%
Since the spin operator [Eq.~(\ref{eq:eq2})] is bilinear in spinons, the transverse  
dipole moment vanishes in the absence of a single--spinon condensate, i.e. 
\bea
\langle S_{\textbf{r}\textbf{r}'}^+ \rangle =0 \quad \text{if} \quad \langle \phi_\textbf{r}\rangle =0 \;.
\eea
Spinons are expected to condense for large value of $|J_\pm|/J_z$, 
in the fully--confined Higgs phase.
However, for intermediate values of coupling, where there is no single--spinon condensate, 
multipole moments can still take on finite values on the bonds of the lattice, without leading 
to a confinement of spinons.

As a concrete example, we consider a quadrupole moment formed 
from transverse spin components, 
%
\bea
\mathcal{\textbf{Q}}=\frac{1}{N}\sum_{\langle ij \rangle}\begin{pmatrix}
\mathcal{Q}_{ij}^1 \\ \mathcal{Q}_{ij}^2 \end{pmatrix}=\langle \frac{1}{N}\sum_{\langle ij\rangle}\begin{pmatrix}
S_i^xS_j^x-S_i^yS_j^y\\ S_i^xS_j^y+S_i^yS_j^x \end{pmatrix}\rangle \;,
\label{eq:nematic}
\eea
where $i, j$ are pyrochlore lattice sites.
Although a finite $\mathcal{\textbf{Q}}$ preserves the time--reversal symmetry for both 
Kramer/non-Kramer materials, it breaks spin--rotation symmetry in the same way as the 
director in a nematic liquid crystal breaks rotation symmetry.
And for this reason states with finite $\mathcal{\textbf{Q}}$ (but vanishing dipole order) are
commonly referred to as a ``spin nematics''.

In terms of diamond--lattice sites, Eq. (\ref{eq:nematic}) can be written
\bea
S_{\textbf{r}\textbf{r}'}^+S_{\textbf{r}'\textbf{r}''}^+=\mathcal{Q}_{ij}^1+i\mathcal{Q}_{ij}^2=\frac{1}{4}\phi_{\textbf{r}'}^\dagger\phi_{\textbf{r}'}^\dagger e^{i(A_{\textbf{r}'\textbf{r}}+A_{\textbf{r}'\textbf{r}''})}\phi_{\textbf{r}}\phi_{\textbf{r}''} \;,
\eea
where the nearest-neighbor diamond sites $\textbf{r}, \textbf{r}'$ ($\textbf{r}', \textbf{r}''$) 
define the pyrochlore site $i$ ($j$). 
A necessary condition for  spin--nematic order is therefore a finite pairing of spinons 
on the bonds of the diamond lattice 
\bea
\Delta_0 \neq 0, \quad \Delta_{\mu-\nu}\neq 0 \quad \text{where} \quad \Delta_{\mu-\nu}=\langle \phi_\textbf{r}\phi_{\textbf{r}+\textbf{e}_\mu-\textbf{e}_\nu}\rangle \;.
\label{eq:nematicpair}
\eea
Such pairing is very naturally motivated by the attractive interactions between spinons 
in Eq. (\ref{eq:Husupp}), and can be studied at mean field level, as described in the main text.
As also noted in the main text, the presence of pair condensate breaks the $U(1)$ gauge 
structure down to $\mathbb{Z}_2$.
And for this reason the spin--nematic phase retains many of the characteristics of a 
$\mathbb{Z}_2$ QSL.

\subsection{Phase transition}

The effective model studied in this Letter
\begin{eqnarray}
\mathcal{H}_\text{rotor}^{''} = \mathcal{H}_{\text{rotor}} + \mathcal{H}_u
\end{eqnarray}
[Eq.~(\ref{eq:H.interacting}) of the main text], contains both terms which mediate both 
the propopagtion of both individual spinons, and pairs of spinons.
For small $|J_\pm|/J_z$ this models support a $U(1)$ QSL, and the Higgs 
transition occurring for $u=0$ [Fig. \ref{fig:schematic}] is already well understood 
in terms of a BEC of spinons, leading to a phase with conventional magnetic 
order \cite{Lee2012}.
We now consider the phase transition occurring as a function 
$u/J_z$, for $J_{\pm}=0$, which occurs through the condensation 
of pairs of spinons.
In this case, the projected rotor model reduces to
\bea
&&\mathcal{H}_{\Delta}\equiv \mathcal{H}_0 + \mathcal{H}_u 
= \frac{J_z}{2}\sum_{\textbf{r}}Q_\textbf{r}^2-u\sum_{\langle\langle \textbf{r}\textbf{r}'' \rangle\rangle}\Big(\phi_{\textbf{r}}^{\dagger}
e^{i(A_{\textbf{r}\textbf{r}'}+A_{\textbf{r}'\textbf{r}''})}\phi_{\textbf{r}''}\Big)^2 +\text{h.c.} \; .
\label{eq:charge-2pro}
\eea
where sum $\langle\langle \textbf{r}\textbf{r}'' \rangle\rangle$ runs over second--neighbour 
bonds of the diamond lattice, with coordination number $z'=2\times \binom 42 = 12$. 
This Hamiltonian is quartic in the spinon fields, and we solve it by seeking a mean--field 
decoupling in terms of suitable order parameters.


The pattern for this calculation follows the well--studied example of single--spinon propagation 
within a rotor model \cite{Senthil2000}.
In this case the term endowing the spinons with dispersion [Eq.~(\ref{eq:H1})] 
can be transcribed in terms of the phases of rotors $\phi_\textbf{r}=e^{-i\varphi_{\textbf{r}}}$, 
to give
\bea
\mathcal{H}_1=-\frac{J_\pm}{2}\sum_{\langle\langle\textbf{r}\textbf{r}''\rangle\rangle}\cos(\varphi_\textbf{r}-\varphi_{\textbf{r}''}+(A_{\textbf{r}\textbf{r}'}+A_{\textbf{r}'\textbf{r}''})),
\eea
and, once a gauge has been fixed, we can anticipate a broken--symmetry state 
\bea
\langle \cos(\varphi_{\textbf{r}})\rangle \neq 0 \quad \text{for} \quad |J_\pm| \rightarrow \infty \; .
\eea
{This phase transition is analogous to the emergence of superconducting order, 
with associated breaking of gauge symmetry.}


By analogy, the second term in Eq. (\ref{eq:charge-2pro}) can be written
\bea
\mathcal{H}_u=-2u\sum_{\langle\langle\textbf{r}\textbf{r}''\rangle\rangle}\cos(2\varphi_\textbf{r}-2\varphi_{\textbf{r}''}+2(A_{\textbf{r}\textbf{r}'}+A_{\textbf{r}'\textbf{r}''})) \;,
\eea
and we anticipate that 
\bea
\langle \cos(2\varphi_{\textbf{r}})\rangle \neq 0  \quad \text{for} \quad u \rightarrow \infty \;.
\eea
where, in the broken symmetry state, the gauge field takes on $\mathbb{Z}_2$
character.   
%
In this phase $u>u_c$, the kinetics of paired spinons $\mathcal{H}_u$ overcomes the energy cost given by $\mathcal{H}_0$ while the single condensate kinetics $\mathcal{H}_1$ fails below the critical value $|J_\pm|<J_c$. As a result, the U(1) gauge fluctuation $ 0 \leq A_{\textbf{r}\textbf{{r}}'} < 2\pi$ is restricted to take a discrete value, either $A_{\textbf{r}\textbf{{r}}'}=0$ or $A_{\textbf{r}\textbf{{r}}'}= \pi$.

In this case, we can fix $(e^{iA_{\textbf{r}\textbf{r}'}})^2=1$ and define an order parameter 
\bea
\Delta=\langle \phi_\textbf{r}\phi_\textbf{r}\rangle 
\eea
so that Eq. (\ref{eq:charge-2pro}) reduces to
\bea
\mathcal{H}_{\Delta}=\frac{J_z}{2}\sum_\textbf{r} Q_\textbf{r}^2-12u\sum_{\textbf{r}}(\Delta\phi_\textbf{r}^{\dagger}\phi_\textbf{r}^{\dagger}+\text{h.c.}-|\Delta|^2)\;, \quad
\label{eq:charge-2MFT}
\eea
The phase of the order parameter $\Delta$ can be eliminated by a further 
gauge transformation $\Delta \rightarrow \Delta e^{2i\Lambda_\textbf{r}}$, so that 
we work with a real, positive $\Delta>0$.


We can now integrate out the gauge charge $Q_\textbf{r}$, and solve 
for the dynamics of spinon pairs in the basis 
\bea
[\phi_{\textbf{k},\omega_n,\alpha}, \phi_{-\textbf{k},-\omega_n,\alpha}^{\dagger}]^T
\eea
where $\alpha \in $ $A$ or $B$ sublattice of diamond lattice sites,  
and $\omega_n =2\pi n T$ is a bosonic Mastubara frequency. 
Doing so, we obtain an (inverse) Green's function
\begin{eqnarray}
&&G^{-1}_{{k_z>0}, \alpha}(i\omega_n)= \left[ 
{\begin{array}{cc}
   \frac{\omega_n^2}{2J_z}+\lambda & -2\cdot 12u\Delta \\
   -2\cdot 12u\Delta &  \frac{\omega_n^2}{2J_z}+\lambda  \\
  \end{array} } 
  \right] \;,
  \nonumber\\
&& G_{k_z>0, \alpha}(i\omega_n)= \frac{1}{(\frac{\omega_n^2}{2J_z}+\lambda +24u\Delta)(\frac{\omega_n^2}{2J_z}+\lambda -24u\Delta)}\left[ 
{\begin{array}{cc}
   \frac{\omega_n^2}{2J_z}+\lambda & 24u\Delta \\
   24u\Delta &  \frac{\omega_n^2}{2J_z}+\lambda  \\
  \end{array} } 
  \right] \;,
\label{eq:invGreenpure}
\end{eqnarray}
where $\lambda$ is the Lagrange multiplier enforcing the rotor 
constraint, cf. Eq.~(\ref{eq:S.rotor.model}).
Here, the additional factor of 2 in the off--diagonal component comes from the 
restriction on momenta
\bea
-12u\sum_{\textbf{k},\alpha= A,B}(\Delta \phi_{\textbf{k},\alpha}^{\dagger}\phi_{-\textbf{k},\alpha}^{\dagger}+\text{h.c.})=-24u\sum_{k_z>0,\alpha}(\Delta \phi_{\textbf{k},\alpha}^{\dagger}\phi_{-\textbf{k},\alpha}^{\dagger}+\text{h.c.})
\;.
\eea
Transforming to imaginary time, and considering the limit of zero temperature 
\bea
G_{\textbf{k},\alpha}(\tau=0)=T\sum_{i\omega_n}G_{\textbf{k},\alpha}(i\omega_n)e^{i0\cdot\omega_n}\overset{T=0\text{K}}{=}\int_{-\infty}^{\infty}\frac{d\omega_n}{2\pi}G_{\textbf{k},\alpha}(i\omega_n)e^{i0\cdot\omega_n } \;,
\eea
we find 
\begin{subequations}
\begin{eqnarray}
G_{\textbf{k},\alpha,11}(\tau =0)=&&G_{\textbf{k},\alpha,22}(\tau=0)=\frac{J_z}{2}(\frac{1}{\omega_1}+\frac{1}{\omega_2}) \;,\\ G_{\textbf{k},\alpha,12}(\tau=0)=&&G_{\textbf{k},\alpha,21}(\tau=0)=\frac{J_z}{2}(\frac{1}{\omega_1}-\frac{1}{\omega_2}) \;,
\end{eqnarray}
\end{subequations}
where the energies of the (localised) spinon bands are given by 
\bea
\omega_{1,2} = \sqrt{2J_z (\lambda \mp 24u \Delta)}
\eea

The values of $\lambda$ and $\Delta$ can then be determined through the self--consistency conditions
\bea
&&\frac{1}{N} \sum_{\textbf{r}\in A}\langle \phi_{\textbf{r}}^\dagger \phi_{\textbf{r}} \rangle = \int_\textbf{k} \langle \phi_{\textbf{k},A}^\dagger \phi_{\textbf{k},A} \rangle (\tau =0) =\int_\textbf{k}G_{\textbf{k},A, 11}=1 \;,
\nonumber\\
&&\frac{1}{N}\sum_{\textbf{r}\in A}\langle \phi_{\textbf{r}}\phi_{\textbf{r}} \rangle = \int_\textbf{k} \langle \phi_{\textbf{k},A}\phi_{\textbf{k},A}\rangle (\tau =0) = \int_{\textbf{k}} G_{\textbf{k},A,21}= \Delta \;,
\label{eq:SEC}
\eea
where $\int_\textbf{k}\equiv \int\frac{d^3k}{V_{\text{BZ}}}$ denotes the $\textbf{k}$-space integral over the Brillouin zone. 
Due to the absence of single spinon propagation, no BEC (of the single spinon) occurs for any value of 
$u/J_z$, and the integrand in Eq.~(\ref{eq:SEC}) is independent of $\textbf{k}$.
It follows that 
\bea
\omega_1=\frac{J_z}{1+\Delta}, 
\quad \omega_2=\frac{J_z}{1-\Delta} \; ,
\label{eq:SECsolve}
\eea
and the ground state energy, Eq.~(\ref{eq:charge-2MFT}), is given by 
\bea
\frac{1}{N}\langle \mathcal{H}_\Delta\rangle = \frac{J_z}{2}\frac{1}{N}\sum_{\textbf{r}}\langle Q_{\textbf{r}}^2\rangle -24u \Delta^2 =\frac{1}{2}\int_\textbf{k}(\omega_1 + \omega_2) -24u \Delta^2 = J_z+(J_z-24u)\Delta^2 +J_z\Delta^4 + \cdots \; .
\label{eq:GSenergy}
\eea
It follows that theres is a 2$^{nd}$--order phase transition at $u_c = J_z/24 \approx 0.0417J_z$,  
between states with $\Delta =0$ and $\Delta \neq 0$.


We conclude by commenting on the order of the phase transition predicted by Eq.~(\ref{eq:GSenergy}).
In Ref.~\citep{Lee2012}, a BEC driven by a similar form of spinon interaction $\sim J_{\pm\pm} S_i^+S_j^+\sim J_{\pm\pm} \phi_{\textbf{r}}^{\dagger}\phi_{\textbf{r}}^{\dagger}\phi_{\textbf{r}+\textbf{e}_\mu}\phi_{\textbf{r}+\textbf{e}_\nu}$ 
was investigated, and found to be first order.
However in this case, the BEC involved a finite single--spinon condensate, $\langle \phi_{\textbf{r}}\rangle \neq 0$.
And within the framework of gauge mean field theory (gMFT), in the limit of weak gauge fluctiations, 
we find that {\it any} BEC driven by spinon interactions, for which $\langle \phi_{\textbf{r}}\rangle \neq 0$, 
will be first order.
Conversly, any continuous transition driven by interactions (in the same limit), 
must necessarily have $\langle \phi_{\textbf{r}}\rangle \equiv 0$, 
consistent with Eq.~(\ref{eq:GSenergy}).
%
%

%
Our argument proceeds as follows.
Suppose that, within a gMFT treatment of a quantum spin ice, a spinon interaction $\mathcal{H}_u^* \sim u(\phi_{\textbf{r}1}^{\dagger}\phi_{\textbf{r}2}^{\dagger}\phi_{\textbf{r}3}\phi_{\textbf{r}4})$ induces a BEC of spinons, $\langle \phi_{\textbf{r}}\rangle \neq 0$, at some $u=u_c$. 
In the condensed phase, if the single spinon condensate $\langle \phi_\textbf{r}\rangle \neq 0$ dominates the physical quantities, we expect $\langle \phi_{\textbf{r}1}\phi_{\textbf{r}2} \rangle \simeq \langle \phi_{\textbf{r}1} \rangle\langle \phi_{\textbf{r}2} \rangle$. 
In this case a mean-field decoupling of the form $\mathcal{H}_u^* \sim u(\phi_{\textbf{r}1}^{\dagger}\phi_{\textbf{r}2}^{\dagger}\langle \phi_{\textbf{r}3}\phi_{\textbf{r}4}\rangle + \langle \phi_{\textbf{r}1}^{\dagger}\phi_{\textbf{r}2}^{\dagger}\rangle \phi_{\textbf{r}3}\phi_{\textbf{r}4}+...)$ is sufficient to describe the BEC transition. 
It follows that, if the BEC at $u=u_c$ is continuous (2$^{nd}$--order) $\langle \phi_\textbf{r}\rangle|_{u=u_c}=0$, 
and the spinon interaction is has no effected at the transition, i.e $\mathcal{H}_{u=u_c}^*\simeq 0$.
At the same time, the BEC should be signalled by an integrable singularity in 
bosonic statistics $\langle \phi_\textbf{k}^{\dagger}\phi_\textbf{k}\rangle$ in $\textbf{k}$-space.
However, at a mean--field level,  $\mathcal{H}_{u=u_c}^* \simeq 0$ cannot lead to any singular
change in the Green function at the phase boundary $u=u_c$. 
A first--order transition, to a state with finite $\langle \phi_{\textbf{r}}\rangle|_{u=u_c} \neq 0$, is therefore 
needed to induce a sudden divergence in the Green function as $u\rightarrow u_c$.
It follows that, within gMFT, any BEC driven by interactions of the form $\mathcal{H}_u^*$
that has a continuous character {\it must} be associated with the onset of an unconventional, ``hidden'', 
phase.
Discontinuous (first--order) transitions may however occur into conventional or unconventional phases.


\section{IV. Phase boundaries}
\label{sec: Phase boundaries}

In Fig.~\ref{fig:ChargeInt} of the main text, we present phase diagram 
exhibiting some of the exotic phases which can be born out of a U(1) QSL.
The characteristics of these phases are listed in Table.~\ref{table:parameter}.   
In this Section, we describe how the phase boundaries shown 
in Fig.~\ref{fig:ChargeInt} were estimated.

\subsection{\textit{Higgs transition}}

The ``Higgs'' phase at large $|J_\pm|/J_z$ is associated with the condensation 
of an individual spinon mode.
This is is signalled by a singularity in the associated spinon Green's function.
We consider spinons moving on a diamond lattice, subject to an XXZ 
model with additional longer--range interactions
\bea 
\mathcal{H}_{\text{XXZ}+} = \mathcal{H}_Q + \mathcal{H}_1 
\; , \qquad
\mathcal{H}_Q = \mathcal{H}_0 +\mathcal{H}_{\text{ZZ}} \; .
\label{eq:H.XXZ+}
\eea
where $\mathcal{H}_0$ is defined through Eq.~(\ref{eq:H0}), $\mathcal{H}_1$ 
through Eq.~(\ref{eq:H1}), and $\mathcal{H}_{\text{ZZ}}$ through Eq.~(\ref{eq:H.ZZ}) of the main text.
The Coulomb interaction term, $\mathcal{H}_Q$, can be transcribed using the Fourier transform,
\bea 
Q_{\textbf{r}\alpha}(\tau)=\frac{1}{\sqrt{N}}\sum_{\textbf{k}} Q_{\textbf{k}\alpha}(\tau) e^{i\textbf{k}\cdot \textbf{r}} \;, 
\eea
where $\alpha = A,B$ is a sublattice index, to give
\bea
\mathcal{H}_Q = \sum_{\textbf{k}}(Q_{\textbf{k}A}^*\;Q_{\textbf{k}B}^*) 
\begin{pmatrix} 
J_z/2 & J_{zz}R_\textbf{k}/2 
\\J_{zz}R_\textbf{k}^*/2 & J_z/2 
\end{pmatrix} 
\begin{pmatrix}
Q_{\textbf{k}A} \\ Q_{\textbf{k}B} 
\end{pmatrix} \;,
\label{eq:Coulomb}
\eea
where 
\bea
R_{\textbf{k}}=\sum_{\mu=0}^3 e^{i\textbf{k}\cdot \textbf{e}_{\mu}} 
\eea
is the structure constant for a diamond lattice defined by vectors $\textbf{e}_{\mu}$.
Above, the additional factor $1/2$ in the off-diagonal term is due to the fact that $Q_{\textbf{r}\alpha}$ is real, i.e. $Q_{\textbf{k}\alpha}=Q_{-\textbf{k}\alpha}^*$.
\bea
\mathcal{H}_{\text{ZZ}} \; &&=\; J_{zz}\sum_{\substack{\la \textbf{r}\textbf{r}'\ra \\ \alpha \in A, \alpha' \in B}} Q_{\textbf{r}\alpha}Q_{\textbf{r}'\alpha'} 
= \frac{J_{zz}}{2} \Big(\sum_{\substack{\la \textbf{r}\textbf{r}'\ra \\ \alpha \in A, \alpha' \in B}} Q_{\textbf{r}\alpha}Q_{\textbf{r}'\alpha'}+\sum_{\substack{\la \textbf{r}\textbf{r}'\ra \\ \alpha \in B, \alpha' \in A}} Q_{\textbf{r}\alpha}Q_{\textbf{r}'\alpha'} \Big)
\nn\\
&& = \; \frac{J_{zz}}{2}\sum_{\textbf{k}} \Big(Q_{\textbf{k}A}^*Q_{\textbf{k}B}\sum_{\mu =0}^3 e^{i\textbf{k}\cdot \textbf{e}_{\mu}} +Q_{\textbf{k}B}^*Q_{\textbf{k}A}\sum_{\mu =0}^3 e^{-i\textbf{k}\cdot \textbf{e}_{\mu}} \Big) \;.
\label{eq:Off-diagonal.Coulomb}
\eea
The coupling of $A$ and $B$ sublattices in Eq.~(\ref{eq:Coulomb}) causes the charge stiffness 
in $\textbf{k}$-space to split into two distinct branches, 
\bea
J_z \rightarrow J_z^{(\pm)} \equiv J_z \pm J_{zz}|R_\textbf{k}| \;,
\label{eq:branch}
\eea
Consistent with this, the first term in the action Eq.~(\ref{eq:Sspinon}) must be modified,  
with the (inverse) charge stiffness $J_z$ in the denominator replaced by  
$J_z$ divided by the determinant of Eq. (\ref{eq:Coulomb}). 
This modifies the phase boundary of Higgs transition associated with spinon condensation.

\begin{figure}[t!]
{\includegraphics[width=0.35\textwidth]{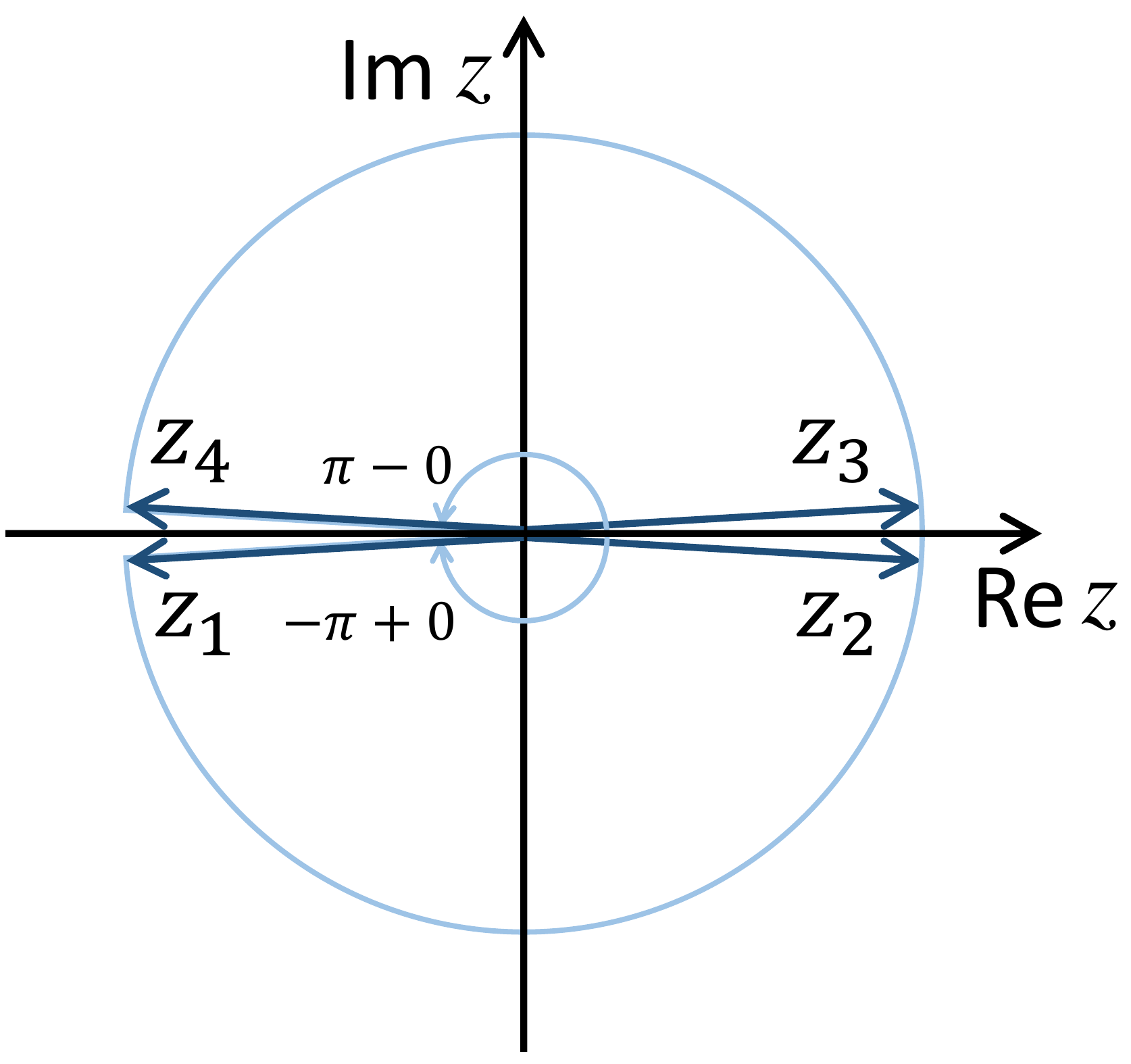}}
\caption{Complex plane on which
4 quantities $z_i$ ($i=1,2,3,4$) are defined. The polar angle of $z=|z|e^{i\vartheta}$ is restricted to be $\vartheta \in [-\pi,\pi)$. All quantities Eq. (\ref{eq:limit}) for $J_z^{(\pm)}>0$ fall into one of $z_1=|z_1|e^{i(-\pi+0)}, z_2=|z_2|e^{i(-0)}, z_3=|z_3|e^{i(+0)}, z_4=|z_4|e^{i(\pi-0)}$.}
\label{fig:CP} 
\end{figure}

By integrating out the gauge charge $Q_{\textbf{r}}$, the (inverse) spinon Green functions of $\mathcal{H}_0+\mathcal{H}_{\text{ZZ}}+\mathcal{H}_1$ are 
\bea
&&G_{\textbf{k}}^{-1}(i\omega_n)=\begin{pmatrix}
\frac{J_z}{2J_z^{(+)}J_z^{(-)}}\omega_n^2 +\lambda - \frac{J_{\pm}}{4}F_\textbf{k} & -\frac{J_{zz}}{J_z^{(+)}J_z^{(-)}}R_\textbf{k} \omega_n^2 \\ -\frac{J_{zz}}{J_z^{(+)}J_z^{(-)}}R_\textbf{k}^* \omega_n^2 & \frac{J_z}{2J_z^{(+)}J_z^{(-)}}\omega_n^2 +\lambda - \frac{J_{\pm}}{4}F_\textbf{k} \end{pmatrix} \;,
\nonumber\\ \nonumber\\
&&G_\textbf{k}(i\omega_n)=\frac{4J_z^{(+)}J_z^{(-)}}{(\omega_n^2+\omega_+^2)(\omega_n^2+\omega_-^2)}\begin{pmatrix}
\frac{J_z}{2J_z^{(+)}J_z^{(-)}}\omega_n^2 +\lambda - \frac{J_{\pm}}{4}F_\textbf{k} & \frac{J_{zz}}{J_z^{(+)}J_z^{(-)}}R_\textbf{k} \omega_n^2 \\ \frac{J_{zz}}{J_z^{(+)}J_z^{(-)}}R_\textbf{k}^* \omega_n^2 & \frac{J_z}{2J_z^{(+)}J_z^{(-)}}\omega_n^2 +\lambda - \frac{J_{\pm}}{4}F_\textbf{k} \end{pmatrix} \;,
\label{eq:QQGreen}
\eea
where $\omega_n$ is a bosonic Mastubara frequency and
$F_\textbf{k}= \sum_{\mu \neq \nu} e^{i\textbf{k}\cdot( \textbf{e}_\mu -\textbf{e}_\nu)}$ describes the spinon propagation. Here, the spinon dispersion $\omega_0^2=2J_z(\lambda -\frac{J_{\pm}}{4}F_\textbf{k})$ is split into two branches 
\bea
\omega_0 \rightarrow \omega_{\pm}=\sqrt{2J_z^{(\pm)}(\lambda -\frac{J_{\pm}}{4}F_\textbf{k})} \;.
\eea
Especially, the diagonal component of Eq. (\ref{eq:QQGreen}) is
\begin{eqnarray}
G_{\textbf{k},11}(i\omega_n)&&=\langle \phi_{\textbf{k}\omega_n A}^*\phi_{\textbf{k}\omega_n A}\rangle =\frac{\frac{J_z}{2J_{z}^{(+)}J_{z}^{(-)}}\omega_n^2 + \lambda - \frac{J_{\pm}}{4}F_\textbf{k}}{(\frac{\omega_n^2}{2J_{z}^{(+)}}+\lambda - \frac{J_{\pm}}{4}F_\textbf{k})(\frac{\omega_n^2}{2J_z^{(-)}}+\lambda - \frac{J_{\pm}}{4}F_\textbf{k})}
\nonumber\\
&&=\frac{1}{2}\Big(\frac{1}{\frac{\omega_n^2}{2J_{z}^{(+)}}+\lambda - \frac{J_{\pm}}{4}F_\textbf{k}} +\frac{1}{\frac{\omega_n^2}{2J_z^{(-)}}+\lambda - \frac{J_{\pm}}{4}F_\textbf{k}}\Big) \;.
\label{eq:QQdiag}
\end{eqnarray}
To evaluate the Mastubara sum $T\sum_{i\omega_n} = \int_{-\infty}^{\infty} \frac{d\omega_n}{2\pi}$ at $T=0$K, we employ, 
\bea
\int_{-\infty}^{\infty}\frac{d\omega_n}{2\pi}\frac{1}{\frac{\omega_n^2}{2J_{z}^{(\pm)}} + \Lambda_\textbf{k}} 
&&= \int_{-\infty}^{\infty}\frac{d\omega_n}{2\pi i} \sqrt{\frac{J_{z}^{(\pm)}}{2\Lambda_\textbf{k}}} ( \frac{1}{\omega_n - i \sqrt{2J_{z}^{(\pm)}\Lambda_\textbf{k}}}-\frac{1}{\omega_n + i \sqrt{2J_{z}^{(\pm)}\Lambda_\textbf{k}}})
\nonumber\\
&&= \frac{1}{2\pi i}\sqrt{\frac{J_{z}^{(\pm)}}{2\Lambda_\textbf{k}}}\text{log}\Big|\frac{\omega_n - i \sqrt{2J_{z}^{(\pm)}\Lambda_\textbf{k}}}{\omega_n + i \sqrt{2J_{z}^{(\pm)}\Lambda_\textbf{k}}}\Big|_{-\infty}^{\infty} = \Theta(J_{z}^{(\pm)})\sqrt{\frac{J_{z}^{(\pm)}}{2\Lambda_\textbf{k}}} \;,
\label{eq:MatInt}
\eea
where $\Theta(x)$ is the usual step function defined in Eq. (\ref{eq:P1}).  In the last step, we have used, 
\bea
  &&\text{In case $J_z^{(\pm)}>0$ } 
  \begin{cases}
    \Big(\omega_n + i \sqrt{2J_{z}^{(\pm)}\Lambda_{\textbf{k}}}\Big)|_{\omega_n\rightarrow \infty} = \Big(\omega_n - i \sqrt{2J_{z}^{(\pm)}\Lambda_{\textbf{k}}}\Big)^*|_{\omega_n\rightarrow \infty}\rightarrow |\omega_n|e^{+i0} \\
	\Big(\omega_n + i \sqrt{2J_{z}^{(\pm)}\Lambda_{\textbf{k}}}\Big)|_{\omega_n\rightarrow -\infty} = \Big(\omega_n - i \sqrt{2J_{z}^{(\pm)}\Lambda_{\textbf{k}}}\Big)^*|_{\omega_n\rightarrow -\infty}\rightarrow |\omega_n| e^{+i\pi}
  \end{cases}
  \nonumber\\
    &&\text{In case $J_z^{(\pm)}<0$ } 
  \begin{cases}
    \Big(\omega_n + i \sqrt{2J_{z}^{(\pm)}\Lambda_{\textbf{k}}}\Big)|_{\omega_n\rightarrow \infty} = \Big(\omega_n - i \sqrt{2J_{z}^{(\pm)}\Lambda_{\textbf{k}}}\Big)^*|_{\omega_n\rightarrow \infty}\rightarrow |\omega_n| \\
	\Big(\omega_n + i \sqrt{2J_{z}^{(\pm)}\Lambda_{\textbf{k}}}\Big)|_{\omega_n\rightarrow -\infty} = \Big(\omega_n - i \sqrt{2J_{z}^{(\pm)}\Lambda_{\textbf{k}}}\Big)^*|_{\omega_n\rightarrow -\infty}\rightarrow -|\omega_n| \;,
  \end{cases}
 \label{eq:limit}
\eea
for any positive real $\Lambda_{\textbf{k}}=\lambda-\frac{J_\pm}{4}F_{\textbf{k}} >0$. In case $J_z^{(\pm)}<0$, Eq. (\ref{eq:limit}) is obvious since all quantities are real. In case $J_z^{(\pm)}>0$, the calculation is illustrated on the complex plane Fig. \ref{fig:CP}.
Then the equal-time Green function of Eq. (\ref{eq:QQdiag}) is
\bea
G_{{\textbf{k}},11}(\tau =0)&&=\Theta(J_z-J_{zz}|R_{\textbf{k}}|)\frac{1}{2}\Big(\sqrt{\frac{J_z - J_{zz}|R_{\textbf{k}}|}{2(\lambda - \frac{J_{\pm}}{4}F_{\textbf{k}})}}+\sqrt{\frac{J_z+J_{zz}|R_{\textbf{k}}|}{2(\lambda - \frac{J_{\pm}}{4}F_{\textbf{k}})}} \Big) 
\nonumber\\
&&+ \Theta(-J_z+J_{zz}|R_{\textbf{k}}|)\frac{1}{2} \sqrt{\frac{J_z+J_{zz}|R_{\textbf{k}}|}{2(\lambda - \frac{J_{\pm}}{4}F_{\textbf{k}})}} \;.
\label{eq:equal-timeG}
\eea
Here, the Lagrange multiplier $\lambda$ is determined by the rotor condition 
\bea
\frac{1}{N}\sum_{{\textbf{r}}\in A}\langle \phi_{\textbf{r}}^*(\tau=0)\phi_{\textbf{r}}(\tau=0)\rangle =\int_{\textbf{k}} G_{{\textbf{k}},11}(\tau =0)= 1 \;.
\label{eq:SECBEC}
\eea
Then the BEC of spinons arises by the integrable divergence in the bosonic statistics  $G_{\textbf{k},11}$. Since $G_{\textbf{k},11}$ should be real
\bea
&&\lambda\geq \frac{J_\pm}{4}F_{\textbf{k}}, \quad 
-4\leq F_{\textbf{k}}\leq 12 \quad\;\; \text{for all} \;\textbf{k}\;,
\eea
from the definition of $F_{\textbf{k}}$. Thus the BEC occurs when $\lambda=3J_{\pm}$ and Eq. (\ref{eq:SECBEC}) becomes
\begin{eqnarray}
J_{\pm}|_{\text{Higgs}}=\Big[\int_{\textbf{k}} &&\Big\lbrace \Theta(J_z-J_{zz}|R_{\textbf{k}}|)\frac{1}{2}\Big(\sqrt{\frac{J_z-J_{zz}|R_{\textbf{k}}|}{2(3 - \frac{1}{4}F_{\textbf{k}})}}+\sqrt{\frac{J_z+J_{zz}|R_{\textbf{k}}|}{2(3 - \frac{1}{4}F_{\textbf{k}})}} \Big) 
\nonumber\\
&&+ \Theta(-J_z+J_{zz}|R_{\textbf{k}}|)\frac{1}{2} \sqrt{\frac{J_z+J_{zz}|R_{\textbf{k}}|}{2(3 - \frac{1}{4}F_{\textbf{k}})}}
\Big\rbrace\Big] ^2 \;.
\label{eq:Higgs}
\end{eqnarray}

\subsection{\textit{Spinon pairing}}

We now consider a model which includes spinon pairing 
\bea
\mathcal{H}  = \mathcal{H}_{\text{XXZ}+} + \mathcal{H}_u
\eea
where $\mathcal{H}_{\text{XXZ}+}$ is defined through Eq.~(\ref{eq:H.XXZ+}), 
and the attractive, on--site, spinon interaction $\mathcal{H}_u$ through 
Eq.~(\ref{eq:H.u}) of the main text.
In the presence of pairing, the Green's function for spinons must be 
enlarged to allow for anomalous off--diagonal terms, as well two sublattices 
\begin{eqnarray}
&&G^{-1}_{k_z>0}(i\omega_n)=\begin{pmatrix} M_{\textbf{k}} & -24u\Delta I_{2\times 2} \\ -24u\Delta I_{2\times 2} & M_{-{\textbf{k}}} \end{pmatrix}  \;,
\nonumber\\
&&M_{\textbf{k}}=M_{-{\textbf{k}}}^*=\begin{pmatrix}
\frac{J_z}{2J_z^{(+)}J_z^{(-)}}\omega_n^2 +\lambda - \frac{J_{\pm}}{4}F_{\textbf{k}} & -\frac{J_{zz}}{J_z^{(+)}J_z^{(-)}}R_{\textbf{k}} \omega_n^2 \\ -\frac{J_{zz}}{J_z^{(+)}J_z^{(-)}}R_{\textbf{k}}^* \omega_n^2 & \frac{J_z}{2J_z^{(+)}J_z^{(-)}}\omega_n^2 +\lambda - \frac{J_{\pm}}{4}F_{\textbf{k}} \end{pmatrix} \qquad \text{[cf. Eq. (\ref{eq:QQGreen})]} \;,
\label{eq:QQpairing}
\end{eqnarray}
in the basis of $(\phi_{\textbf{k},\omega_n,A},\phi_{\textbf{k},\omega_n,B},\phi_{-\textbf{k},-\omega_n,A}^{\dagger},\phi_{-\textbf{k},-\omega_n,B}^{\dagger})^T$ and $I_{2\times 2}$ is the $2\times 2$ identity matrix. For the unitary matrix $U_{\textbf{k}}$ diagonalizing, 
\bea
U^{\dagger}_{\textbf{k}}M_{\textbf{k}}U_{\textbf{k}}=U_{\textbf{k}}^TM_{-\textbf{k}}U_{\textbf{k}}^*=D_{\textbf{k}} \;,
\label{eq:Uk}
\eea
we rearrange Eq. (\ref{eq:QQpairing}),
\begin{eqnarray}
\begin{pmatrix}
U_\textbf{k}^{\dagger} & 0 \\ 0 & U_\textbf{k}^T \end{pmatrix} 
\begin{pmatrix}
M_\textbf{k} & -24u\Delta I_{2\times 2} \\ -24u\Delta I_{2\times 2} & M_{-\textbf{k}} \end{pmatrix} 
&&\begin{pmatrix}
U_\textbf{k} & 0 \\ 0 & U_\textbf{k}^* \end{pmatrix} =
\begin{pmatrix}
D_\textbf{k} & -24u\Delta U_\textbf{k}^{\dagger}U_\textbf{k}^* \\-24u\Delta U_\textbf{k}^TU_\textbf{k} & D_\textbf{k} \end{pmatrix} \cong
\begin{pmatrix}
D_\textbf{k} & -24u\Delta I_{2\times 2} \\ -24u\Delta I_{2\times 2} & D_\textbf{k} \end{pmatrix} ,
\nonumber\\
D_\textbf{k}=&&\begin{pmatrix}
\frac{\omega_n^2}{2J_z^{(+)}}+(\lambda-\frac{J_{\pm}}{4}F_\textbf{k}) & 0 \\ 0 & \frac{\omega_n^2}{2J_z^{(-)}}+(\lambda-\frac{J_{\pm}}{4}F_{\textbf{k}}) \end{pmatrix} .
\label{eq:QQdiagpairing}
\end{eqnarray} 
Since the Green function is influential near the origin $\textbf{k}=(0,0,0)$, the unitary matrices $U_k$ and $M_k$ are almost real, $U_{\textbf{k}}^{\dagger}U_{\textbf{k}}^* \approx I_{2\times 2}$ which is harmless for small $u\Delta$. As a result, Eq. (\ref{eq:QQpairing}) is decoupled into 2 matrices.
\begin{eqnarray}
G^{-1}_{k_z>0}(i\omega_n)
=\begin{pmatrix}
M_{\textbf{k}}^+ & 0 \\ 0 & M_\textbf{k}^-   
\end{pmatrix}
\quad \text{where} \quad M_\textbf{k}^{\pm}=\begin{pmatrix}
\frac{\omega_n^2}{2J_{z}^{(\pm)}}+(\lambda -\frac{J_{\pm}}{4}F_\textbf{k})  & -24u\Delta \\ -24u\Delta & \frac{\omega_n^2}{2J_{z}^{(\pm)}}+(\lambda -\frac{J_{\pm}}{4}F_\textbf{k})  
\end{pmatrix} \;.
\end{eqnarray} 
Likewise Eqs. (\ref{eq:QQdiag})-(\ref{eq:equal-timeG}), the equal-time Green functions are
\begin{eqnarray}
G_{\textbf{k},11}&&=G_{\textbf{k},22}=\frac{1}{2}\Big(\sqrt{\frac{J_z^{(+)}}{2(\lambda - \frac{J_{\pm}}{4}F_\textbf{k}-24u
\Delta)}}+\sqrt{\frac{J_z^{(+)}}{2(\lambda - \frac{J_{\pm}}{4}F_\textbf{k}+24u\Delta )}} \Big) \;,
\nonumber\\
G_{\textbf{k},12}&&=G_{\textbf{k},21}^*=\frac{1}{2}\Big(\sqrt{\frac{J_z^{(+)}}{2(\lambda - \frac{J_{\pm}}{4}F_\textbf{k}-24u
\Delta)}}-\sqrt{\frac{J_z^{(+)}}{2(\lambda - \frac{J_{\pm}}{4}F_\textbf{k}+24u\Delta )}} \Big) \;,
\nonumber\\
G_{\textbf{k},33}&&=G_{\textbf{k},44}=\Theta(J_z^{(-)})\frac{1}{2}\Big(\sqrt{\frac{J_z^{(-)}}{2(\lambda - \frac{J_{\pm}}{4}F_\textbf{k}-24u
\Delta)}}+\sqrt{\frac{J_z^{(-)}}{2(\lambda - \frac{J_{\pm}}{4}F_\textbf{k}+24u\Delta )}} \Big) \;,
\nonumber\\
G_{\textbf{k},34}&&=G_{\textbf{k},43}^*=\Theta(J_z^{(-)})\frac{1}{2}\Big(\sqrt{\frac{J_z^{(-)}}{2(\lambda - \frac{J_{\pm}}{4}F_\textbf{k}-24u
\Delta)}}-\sqrt{\frac{J_z^{(-)}}{2(\lambda - \frac{J_{\pm}}{4}F_\textbf{k}+24u\Delta )}} \Big) \;.
\label{eq:equal-timeGpairing}
\end{eqnarray}
The pairing instability out of U(1) QSL is evaluated  by self-consistent equations of $\lambda, \Delta\neq 0$.
\begin{eqnarray}
&&\frac{1}{N}\sum_{\textbf{r}\in A, B}\langle\phi_{\textbf{r}}^*\phi_{\textbf{r}} \rangle =\int_{\textbf{k}} (G_{\textbf{k},11}+G_{\textbf{k},33}) =2 \;,
\nonumber\\
&&\frac{\partial}{\partial \Delta} \langle \mathcal{H}_{\text{rotor}}'\rangle = 24u\Big(-\int_\textbf{k} (G_{\textbf{k},12}+G_{\textbf{k},34})+2\Delta \Big)=0 \;.
\end{eqnarray}

\subsection{\textit{Gauge charge polarization}}
In case $J_{\pm} =0$, the ground state energy per unit cell is simply
\bea
E[\delta Q]|_{J_{\pm}=0}=2\cdot\frac{J_z}{2}(\delta Q)^2 -4 J_{zz}(\delta Q)^2 \;,
\label{eq:QQE1}
\eea 
where the coefficients $2$ and $4$ count the number of sublattices and bonds connecting them respectively. The phase transition between $\delta Q =0 $ and $\delta Q \neq 0$ is strongly 1$^{st}$--order since $\delta Q =0 $ for $J_{zz}<J_z/4$ and $\delta Q = \delta Q|_{\text{max}}=2$ for $J_{zz}>J_z/4$.

\begin{figure}[]
{\includegraphics[width=0.6\textwidth]{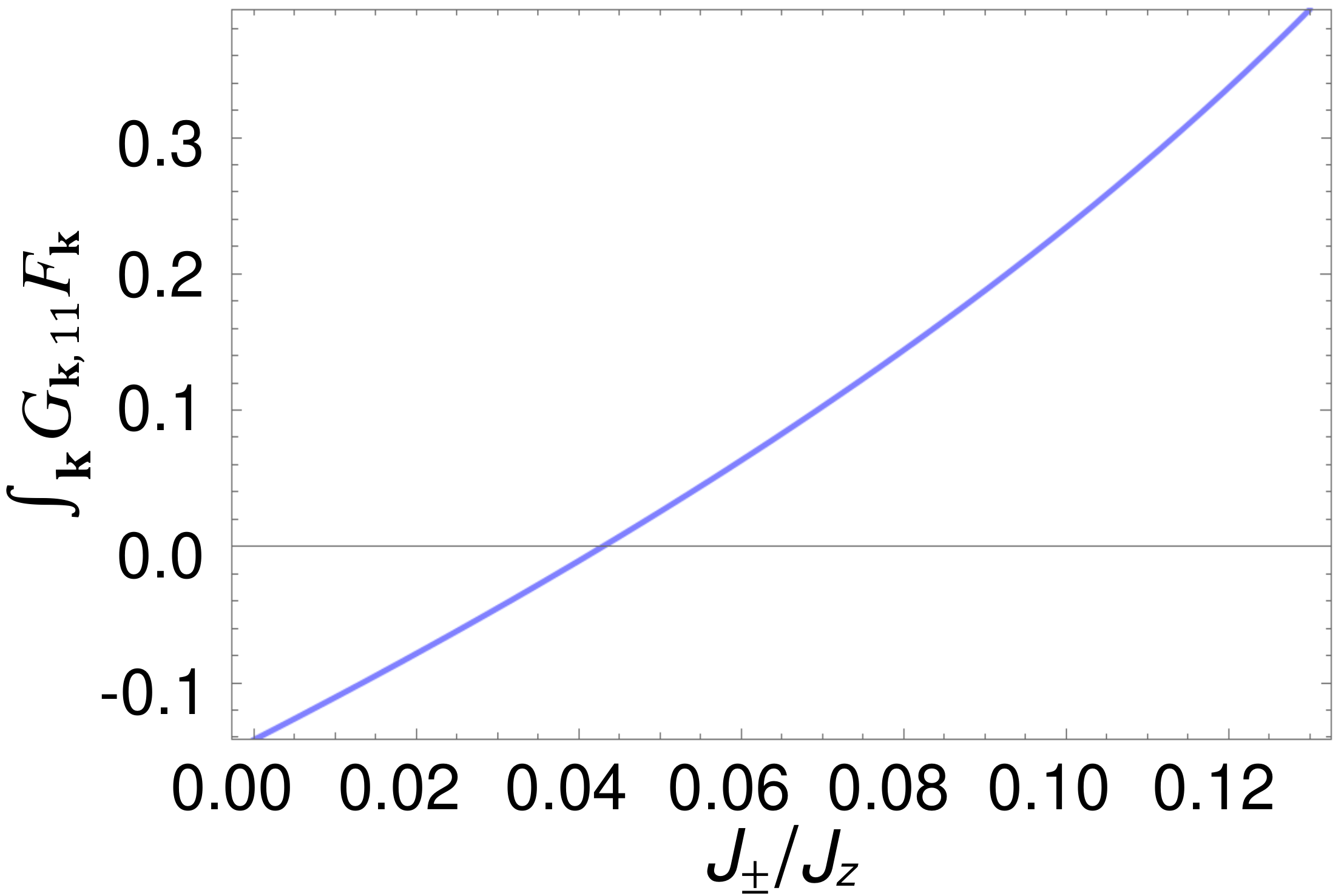}}
\caption{Plot of $\int_{\textbf{k}}G_{\textbf{k},11}F_\textbf{k} $ versus $J_\pm/J_z$ at fixed $J_{zz}/J_z=0.25$, which flips its sign at $J_\pm/J_z=0.043$.} 
\label{fig:Greenplot}
\end{figure}

Turning on the spinon kinetics $J_{\pm}>0$, we desire the fourth terms $\sim(\delta Q)^4$ in $\langle \mathcal{H}_1+\delta H_{\text{rotor}}\rangle$ for a partial polarization $0<\delta Q < \delta Q|_{\text{max}}$ which allows the in-plane magnetization or the spinon pairing marginally. 
In the static limit with $\delta Q(i\omega =0)$, 
$\langle \mathcal{H}_1\rangle$ hardly loosen the full polarization $\delta Q|_{\text{max}}$ since the spinon and gauge charge fields are decoupled.
Then we consider only the correction $\langle \delta\mathcal{H}_{\text{rotor}}\rangle$ and fixing $s_{\textbf{r}\textbf{r}'}^+=1/2$ for simplicity. 
Since Eq. (\ref{eq:QQE1}) is the classical mean field energy of $\delta Q$, the decoupled product evaluation $\langle f[Q_{\textbf{r}}]\phi_{\textbf{r}}^{\dagger}\phi_{\textbf{r}''} \rangle \approx f[Q_{\textbf{r}\in A}=-Q_{\textbf{r}\in B}=\delta Q]\langle \phi_{\textbf{r}}^{\dagger}\phi_{\textbf{r}''}\rangle$ of Eq. (\ref{eq:deltaHXXZ}) is a plausible approximation.

\bea
\frac{1}{N}\langle \delta \mathcal{H}_{\text{rotor}}\rangle \approx -\frac{J_\pm}{4}\Big(\frac{7}{90}\delta Q^2 -\frac{78689}{1440000}\delta Q^4+...   \Big)\Big(\int_\textbf{k} G_{\textbf{k},11}F_\textbf{k}\Big) \;,
\label{eq:appxdeltaH}
\eea
Then the quadratic term $\sim (\delta Q)^2$ corrects the critical value $J_{zz}|_{c,(J_\pm=0)}=J_z/4$.
\bea
J_{zz}|_c \approx \frac{J_z}{4}-\frac{7J_{\pm}}{1440}\Big(\int_\textbf{k} G_{\textbf{k},11}F_\textbf{k}\Big) \;.
\label{eq:criJzz}
\eea
and the quartic correction $\sim (\delta Q)^4$ smooths out the phase transition boundary of Eq. (\ref{eq:QQE1}). As long as the spinon propagation $\langle \phi_\textbf{r}^{\dagger}\phi_{\textbf{r}''}\rangle <0$ is negative for $J_\pm/J_z<0.043$, the phase transition is the 1$^{st}$--order whereas the positive spinon propagation allows the 2$^{nd}$--order transition for $J_\pm/J_z>0.043$. (Fig. \ref{fig:Greenplot})

\end{document}